\documentclass[compsoc,conference,a4paper,10pt,times]{IEEEtran}
\IEEEoverridecommandlockouts

\usepackage{tikz}
\usepackage{amsmath}
\usepackage{epsfig}

\usepackage{cite}
\usepackage{textcomp}
\usepackage{bmpsize}
\usepackage{lipsum}
\usepackage[colorlinks=true,urlcolor=black]{hyperref}
\def\BibTeX{{\rm B\kern-.05em{\sc i\kern-.025em b}\kern-.08em
    T\kern-.1667em\lower.7ex\hbox{E}\kern-.125emX}}

\usepackage[utf8]{inputenc}
\usepackage{epsfig, hyperref, comment, amsmath}
\usepackage{color}
\usepackage{subfig}
\usepackage{xspace}
\usepackage{xcolor}
\usepackage{multirow}
\usepackage{empheq}
\usepackage{tabularx}
\usepackage{graphicx}
\usepackage{booktabs}

\usepackage{tikz}
\usetikzlibrary{shapes,arrows}
\tikzset{font=\scriptsize}

\usepackage{algorithm}% http://ctan.org/pkg/algorithms
\usepackage{algorithmicx}
\usepackage{algpseudocode}
\usepackage{colortbl}

\renewcommand{\algorithmicrequire}{\textbf{Input:}}

\newcommand{\authnote}[2]{{\bf \textcolor{blue}{#1}: \em \textcolor{red}{#2}}}

\newcommand{\desmond}[1]{\authnote{Desmond}{#1}}

\newcommand{\name}{\textsf{SIERRA}\xspace}

\makeatletter
\newcommand{\linebreakand}{%
  \end{@IEEEauthorhalign}
  \hfill\mbox{}\par
  \mbox{}\hfill\begin{@IEEEauthorhalign}
}
\makeatother

\begin{document}

\title{\name: Ranking Anomalous Activities in Enterprise Networks}

\author{
\IEEEauthorblockN{Jehyun Lee}
\IEEEauthorblockA{Trustwave\\
Jehyun.Lee@trustwave.com}
\and
\IEEEauthorblockN{Farren Tang}
\IEEEauthorblockA{Trustwave\\Farren.Tang@trustwave.com}
\and
\IEEEauthorblockN{Phyo May Thet\IEEEauthorrefmark{1}}

\IEEEauthorblockA{A*STAR $\text{I}^2\text{R}$\\Phyo\_May\_Thet@i2r.a-star.edu.sg}
\linebreakand 
\IEEEauthorblockN{Desmond Yeoh\IEEEauthorrefmark{2}}

\IEEEauthorblockA{Shopee\\desmond.yeohtb@shopee.com}
\and
\IEEEauthorblockN{Mitch Rybczynski}
\IEEEauthorblockA{Trustwave\\mrybczynski@trustwave.com}
\and
\IEEEauthorblockN{Dinil Mon Divakaran}
\IEEEauthorblockA{Trustwave\\Dinil.Divakaran@trustwave.com}
}

\maketitle
\begingroup\renewcommand\thefootnote{\IEEEauthorrefmark{1}}
\footnotetext{The work was done when the author was with Trustwave.}
\endgroup
\begingroup\renewcommand\thefootnote{\IEEEauthorrefmark{2}}
\footnotetext{The work was done when the author was associated with NUS-Singtel Cyber Security R\&D Corp. Lab.}
\endgroup

\begin{abstract}
An enterprise today deploys multiple security middleboxes such as firewalls, IDS, IPS, etc. in its network to collect different kinds of events related to threats and attacks. These events are streamed into a SIEM (Security Information and Event Management) system for analysts to investigate and respond quickly with appropriate actions. However, the number of events collected for a single enterprise can easily run into hundreds of thousands per day, much more than what analysts can investigate under a given budget constraint (time). In this work, we look into the problem of prioritizing suspicious events or anomalies to analysts for further investigation. We develop \name, a system that processes event logs from multiple and diverse middleboxes to detect and rank anomalous activities. \name takes an unsupervised approach and therefore has no dependence on ground truth data. Different from other works, \name defines {\em contexts}, that help it to provide visual explanations of highly-ranked anomalous points to analysts, despite employing unsupervised models. We evaluate \name using months of logs from multiple security middleboxes of an enterprise network. The evaluations demonstrate the capability of \name to detect top anomalies in a network while outperforming naive application of existing anomaly detection algorithms as well as a state-of-the-art SIEM-based anomaly detection solution. 
\end{abstract}


\section{Introduction}

Security Information and Event Management, or SIEM, is a widely used security system for enterprises~\cite{marqual2020global}. SIEM gathers events in logs from various security middleboxes (MBs) such as firewall, IDS (intrusion detection system), IPS (intrusion prevention system), web-proxy, host OS audit systems\footnote{Although host audit systems are not strictly middleboxes, our use of the term `middleboxes' includes them for simplicity.}, etc. Thus a SIEM has visibility of suspicious activities in an enterprise, which allows it to analyze and detect different forms of threats and attacks facing the enterprise. 

One of the challenges in analyzing SIEM logs is the sheer volume of the events that it collects on a regular basis~\cite{yen2013beehive,oprea2018made,hassan2020tactical}.  Processing these humongous number of events coming from multiple MBs is impractical, not only because of the cost of analysis but also due to the fatigue it causes to human analysts~\cite{coresecurity2021siem}. Therefore, as also recognized by the industry, there is a need to prioritize important (set of) events for analysts that in turn helps them to prioritize their investigation.  High priority events can be detected by a supervised model trained on `high priority' and `low priority' labeled events. However, labeling all events streamed to a SIEM is both labour-intensive and unsustainable in the long run; and supervised models cannot be kept up-to-date without continuous availability of labeled datasets.  We identify the following challenges in detecting high priority or {\em anomalous} events in a SIEM environment.\\

\noindent \textbf{Diversity of middleboxes and their log semantics.} In contrast to a series of simple values, like  temperature readings from a sensor or  count of signals, the events from a security MB have complex and diverse semantics. Security MBs are distinct in their functionality and deployed locations; they are also developed by several different vendors. These differing aspects also manifest in their logs. For example, a firewall of one vendor may specify the CVEs related to some events while logging, whereas one from another vendor might not. Similarly, a firewall logs information related to traffic, and this is  typically based on the policy of the network admin. For example, some administrators might choose to log even accepted and normal connections along with rejected/dropped connections, while others might log only the latter. Besides, it is also important to note that, all the events logged by a firewall are not related to serious security incidents. On the other hand, a DPI (deep packet inspection) or IDS system would often log events when the packets or traffic flows match any of the pre-defined rules. This means, DPI and IDS events are likely more suspicious than firewall events. In short, semantics of logs of different MBs are different.\\

\noindent \textbf{Diversity of target environments.} A SIEM system collects events from numerous enterprise networks (customers). These networks are different in the sizes (in terms of number of users, network capacity, etc.), in the types of computing devices connected, the security MBs deployed, the logging configurations for MBs, etc. This essentially means, to be of practical use for a SIEM that collects and analyzes logs from multiple enterprise networks, a system to detect anomalous activities in one enterprise network should be able to adapt automatically to another enterprise network without manual fine-tuning and labeling. \\

\noindent \textbf{Ground truth limitation.} Due to large volumes of events streaming into a SIEM, it is simply not possible to generate ground truth for all the events. Due to the allocated and limited budget (time), analysts typically  analyze only a few events to verify whether there are serious security incidents; and this means, verified information is only available for a fraction of events reaching a SIEM. Therefore, any solution for detecting anomalies in a SIEM should not be depending on datasets of labeled information.\\

Most of the previous anomaly detection studies in this context fall short in one or more of these challenges.  The solutions relying on specific types of attacks (e.g., phishing, malware infection, and APT)~\cite{d-fence-2021, phishpedia-2021,  oprea2018made,hassan2020tactical,najafi2019malrank} can be applied to only those networks that deploy specific types of MBs (e.g., email gateway, web proxy, DNS server, IDS, and DHCP server)  providing the dependent data (e.g., URLs, domain names, specific intrusions, etc.).  Besides, often such solutions are also dependent on labeled data~\cite{farshchi2015experience,zhang2016automated,he2016experience} (including data from emulated environment, which is limited in the representation of real-world anomalies).

To overcome the above challenges, we propose \name (\underline{S}ecurity \underline{I}nformation and \underline{E}vent \underline{R}isk \underline{R}ank \underline{A}nalysis system), a frontline solution for SIEM event logs, that detects and provides high ranked anomalies  to SOC (security operations center) analysts for further investigation. \name incorporates multiple novel aspects: i)~Different from past works, \name employs an unsupervised approach to detect anomalous activities (set of events) with no hard constraints for practical deployment in different enterprise networks; in other words, \name does not depend on synthetic dataset, availability of ground truth, or manual fine-tuning. ii)~\name defines and employs numerous features that capture the security conditions observed by diverse MBs. iii)~To the best of our knowledge, \name is the first solution that defines {\em contexts} for capturing semantics pertaining to activities of an enterprise network (Section~\ref{sec:CF}); and this makes \name effective in handling multiple networking contexts among different MBs. iv)~With the ability to identify the contexts and the features that lead the anomalous points to be highly ranked, \name provides analysts with visual explanations (as time-series) useful for investigations (see Section~\ref{sec:scoring:case}).

We evaluate \name using nineteen weeks across five months of real SIEM data from multiple MBs of an enterprise network (see Section~\ref{sec:evalution} for details). First, we compare \name with the state-of-the-art SIEM-based anomaly detection solution, namely DeepLog~\cite{du2017deeplog}. Second, since our design of \name is such that it can work with different one-class models (e.g., Isolation Forest, OCSVM, Variational Autoencoder, etc.), we also evaluate \name with multiple such models. Furthermore, we carry out experiments to study the efficacy of our specifically defined techniques (i)~to deal with varying semantics of logs of different security MBs, and (ii) to represent rich contexts of network information. Based on these extensive experiments, we observe that \name is capable of detecting top anomalies in a network, aided significantly by our novel contextual information (captured as features). In particular, \name outperforms the naive application of existing anomaly detection algorithms as well as DeepLog. 

After presenting the overview of \name in Section~\ref{sec:system}, we design and develop \name in three sections. In Section~\ref{sec:CF}, we first define the event contexts and the contextual-features that capture the anomalies, and subsequently describe the modeling phase and the analysis phase of \name in Section~\ref{sec:modeling} and Section~\ref{sec:scoring}, respectively.  In Section~\ref{sec:evalution}, we evaluate the performance of \name with a real-world enterprise SIEM log dataset. We demonstrate how \name supports a security analyst with an investigation of high-ranked anomalies detected by \name in Section~\ref{sec:scoring:case}.

\section{Motivation and Problem Definition}\label{sec:motivation}

\subsection{Motivation}

The problem we address exists because middleboxes, such as firewall, IDS, OS audit systems, etc., generate numerous events, much of which are irrelevant or unrelated to a serious threat or attack; i.e., they are neither accurate nor precise. For example, an IDS or a firewall often has a warning rule with (and for) only a specific destination port number. Clearly, such a generic rule generates numerous events, which on their own, are irrelevant and ignored.
Indeed, in our dataset, we have an IDS generating events for every resolution of *.tw domain. Clearly, this is too general a rule that creates many events at an analyst’s end.

This generation of high-volume events happens not only due to wrong or obsolete rules (resulting in false positives and false negatives), lack of specific information for rule generation (resulting in false negatives), etc., but also due to the disconnect between an administrator of an enterprise network and a security analyst (often based at a SOC). An administrator’s role is to log as many suspicious activities as possible, and therefore it is only natural that an administrator's policy is general in nature. On the other hand, an analyst is the one who investigates the generated events for detecting threats and attacks, to report back potential security incidents and their severity. This is also how SIEM generally works, and the reason why it exists --- enterprises ingest high-volume logs into SIEM for analysts to figure out the security incidents.

The goal of this work is to mine statistical and correlated anomalies from voluminous events generated by multiple MBs of an enterprise network. To understand the relationship between a SIEM  anomaly detection solution and MBs, consider the following use case. The events of incoming scanning attempts logged by a firewall at an enterprise are not interesting to an analyst (as they are very common). Yet unusual scanning patterns need to be investigated, for example, an {\em outgoing} scanning activity that has temporal correlation with events from other MBs (e.g., IDS) due to a new attack vector. Indeed, that the scanning activity is {\em outgoing}, i.e., it is from an internal host to an external host, in itself provides a {\em context} that brings more attention to the event (it is likely an internal host that is scanning is infected).  We argue that an anomaly detection solution should detect and provide the context for such investigation. Note that, our goal here is not to improve any middlebox specifically (in which case, raw events such as network traffic should be processed), but instead to analyze SIEM events and thereby make it easier for an analyst to investigate further. In other words, we aim to develop a solution for SIEM that collects large-scale events from MBs of enterprises, which have to be prioritized for effective and efficient investigation.

\subsection{Problem definition}
Based on the above challenges, we argue that a SIEM-based anomaly detection solution should target to achieve the following:

\begin{itemize}
\item \textbf{Anomaly modeling with minimal event semantics:}
the features used to define an anomaly detection solution should be general enough to capture the semantics across different security middleboxes and their various event types. Put differently, the semantics being extracted should not be highly dependently on any particular MB. \\

\item \textbf{Middlebox and network agnostic modeling:}
the anomaly detection system should capture the security conditions by itself from the event logs generated by any security middlebox usually deployed in enterprise networks. \\

\item \textbf{Attack type independent detection:} the anomaly detection system should detect anomalous behaviors even if the behavior is not pre-defined or known as a specific type of attack, as far as the behavior is observed in any form of security events by a SIEM system.
\end{itemize}

\noindent \textbf{Assumption.} We assume that the logs collected by a SIEM system are in the parsed form with the commonly expected fields, e.g., timestamp and network flow information. We do not consider the problem of inconsistent and unknown log format. 

\section{\name: Overview}\label{sec:system}

\begin{figure}[t]
    \centering
    \includegraphics[width=0.48\textwidth]{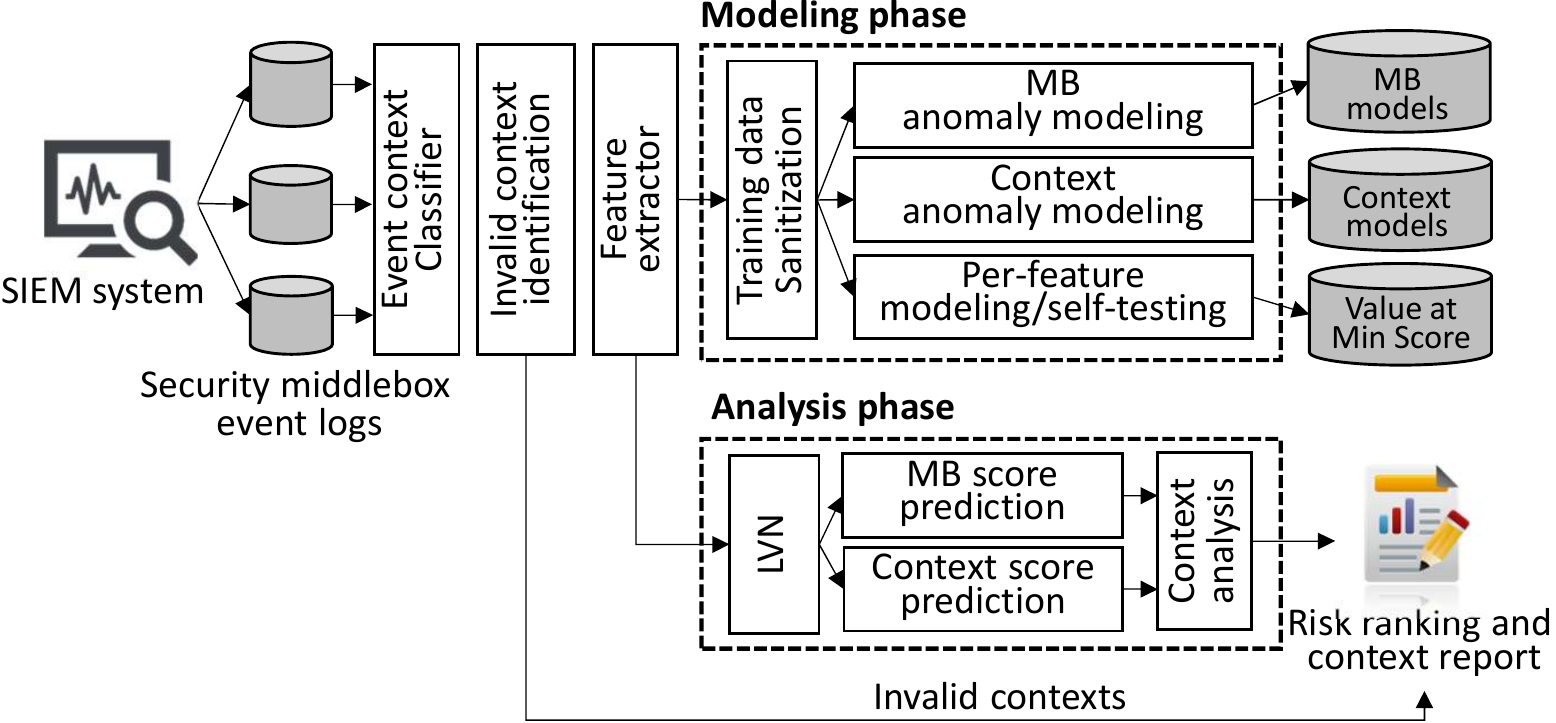}
    \caption{\name system architecture}
    \label{fig:overview}
\end{figure}

Fig.~\ref{fig:overview} presents the overall architecture of \name.  \name defines and uses  contextual features (CF) from the SIEM event logs, for quantifying and modeling security conditions of a given enterprise network. The events from different MBs (middleboxes) collected at the SIEM form the input to \name. \textit{Event context classifier} (Section~\ref{sec:CF:context}) identifies the context of each event  based on the semantic that is defined by its corresponding security MB, network flow direction  triggering the event, and the role of the internal host(s) in the particular event.  We provide the details of context definition in Section~\ref{sec:CF}. For the classified events, \textit{Feature extractor} (Section~\ref{sec:CF:FE}) extracts statistical information representing security conditions (e.g., the number of unique source IP addresses) from the set of events in each context. 

In \textbf{Modeling phase}, \name first sanitizes the training dataset by employing inter-quantile range (IQR) method to detect and filter outliers. Then it employs multiple unsupervised modeling techniques for detecting and scoring anomalies. 
The three different types of anomaly detection models trained in this phase are: MB model, Context model, and Per-feature model. \name later uses MB model and Context model in the \textbf{Analysis phase}, the former for anomaly scoring and the latter for explaining the anomalies. We describe these models later in Section~\ref{sec:modeling}. 

\textbf{Analysis phase} first performs low-value normalization (LVN) to the test feature values, and then detects anomalies using the MB models. This gives us MB-level anomaly scores for a testing time-slot. Risk scoring and ranking process measures the global anomaly score of a given time window by aggregating the individual MB's anomaly rankings for the time-slot. We present this scoring and aggregation processes in Section~\ref{sec:scoring}. 

\section{Contextual features}\label{sec:CF}
We now present the motivation and definition of the contextual features. The features in \name are computed in fixed length intervals. For example, every 10 minute, we extract the statistics of the events, such as `number of unique source IP addresses', `number of unique source ports per unique destination port', etc. The contextual features represent these statistics along with the context of the events.

\subsection{Event context classification}\label{sec:CF:context}
Three properties of an event are used to define its context: i)~the general severity level of an event, ii)~the network flow direction, and iii)~the type of action on an event. Rationale of the context properties is based on the high-level description of security events, that  are common in most of existing MBs. We list the possible and general values for these contextual properties in Table~\ref{tab:context_prop}.

\noindent (i) \underline{Event severity}: The event severity differentiates the information level events from security events. We define two levels of event severity. One set of events are those that are used only for notification purposes and are not of  any security concern. The other set of events are warnings and alerts that are generated due to suspicious or insecure activities. As we argue in our motivation, we define the contextual values as general as possible so as to apply them to a wide range of security middleboxes without being constrained by their vendor, policy definitions, and MB-specific semantics.  Therefore, we define only two classes of event severity, though a security MB may have more fine-grained definitions.\\

\noindent (ii) \underline{Network flow direction}: When a network security MB such as an IDS or a firewall generates an event, the event log usually provides the source and destination IP addresses (or hostnames) and network ports.  Depending on the deployment point of the security MB, one connecting host in an event must be from within the internal network, i.e., LAN or DMZ. The other end can be in the LAN, DMZ or WAN (Internet).  The events with no end-point address, a loopback address or broadcast are considered as Others, and the erroneous cases, e.g., a flow from LAN-side interface but with invalid IP address for LAN, are the exceptions.  As most of the security policies (e.g., firewall rules and IDS signatures) provide the direction of the network flows (i.e., incoming or outgoing), the event direction is a commonly available contextual information for further investigation and can assist in prioritizing events for analysis. \\

\noindent (iii) \underline{Type of action}: The {\em action} context helps in determining whether a host in the internal network of an enterprise is a potential victim or an adversary.  Together with the context of network flow direction, with the action of the involved internal host(s), we get more information about the threat.  For example, an event related to an intrusion attempt to an internal server mostly generates a log in its \textit{Request} action, but an event corresponding to \textit{Respond} from an internal server indicates possible information leakage. When both end-points are internal, we consider the action of the source host as the action type for the event.  Lastly, the  action not related to network are considered as `Others'. The list of actions in Table~\ref{tab:context_prop} are considering the network middleboxes; however this can be extended to cover host-based actions. 
\begin{table}[t]
\centering
\caption{List of event context properties and values}
{
\begin{tabular}{ll@{}c} \toprule
Property & & Value \\ \toprule
Event severity & & \{ \texttt{Info,Warn} \} \\ \hline
\multirow{2}{*}{Network flow} & Src.& \{ \texttt{LAN,WAN,DMZ,Others,Exception} \} \\
 & Dst.& \{ \texttt{LAN,WAN,DMZ,Others,Exception} \} \\  \hline
Type of action & & \{ \texttt{Request,Response,Others} \} \\ \bottomrule
\end{tabular}\label{tab:context_prop}
}
\end{table}

\subsection{Invalid context identification}

As motivated in Section~\ref{sec:motivation}, MBs have distinct logging behavior because of its topological position, functionality, logging policies, and services in operation. 
For example, the logs from $\text{FW}_{1}$ do not have events for the LAN-to-LAN and WAN-to-LAN traffic. Thus occurrence of LAN-to-LAN events are anomalous based on the historical information.  
However, such information is not readily made available to a detection solution, and it is a challenge to learn them automatically. 
\name learns this validity of event contexts from the given dataset and utilizes them efficiently to detect the occurrence of unexpected events. Once \name builds a list of valid contexts, the events not in the valid contexts are immediately considered as an anomaly.

Avoiding the events in an invalid context significantly reduces the feature dimension compared to the single multi-feature MB model.
In $\text{FW}_{1}$, $\text{IDS}_{1}$ and $\text{FW}_{2}$, there are 74, 73 and 98 number of contexts, respectively, that are not observed, while there are a total of 104 possible contexts. The infeasible contexts are mostly because of the visibility of each MBs on network flows, depending on the topological position of the MBs, and monitoring policies.

\subsection{Feature definition}\label{sec:CF:FE}
For the events in each context, we define features that measure the statistics representing the frequency, diversity and distribution of security events; they are listed in Table~\ref{tab:feature_list}.  \name has 6 simple count-based features, and 10 rate-based features. \name emphasize the changes in the main actors with 2 features for top-10 source and destination hosts.  Changes in 12 well-known service ports, e.g., HTTP, HTTPS, SSH, DNS, etc, have 3 features per port, making a total 36 features. Lastly, biased behavior by a few highly active IP addresses or ports are captured by 8 Gini-index features. Features are computed at every fixed-length interval from the SIEM events generated during that interval. 
The features are extracted for each context. For example, while a simplest form extracts the feature `\texttt{number of unique source ports}' from all event logs of an MB within a time-slot (e.g. 10 minutes), the feature with the context \texttt{[Warn,LAN,WAN,Request]} is extracted only from the \textit{warning} events which satisfy the condition `when a source host of the corresponding network flow is in LAN, and the destination host is a server in WAN'.

\begin{table}[t]
\centering
\caption{\name Feature list}
{
\begin{tabular}{ll} \hline
 Feature category & Feature \\ \hline
 & Num. of events \\ 
 & Uniq. src IP addrs \\
 Count & Uniq. dst IP addrs \\
 (6 features) & Uniq. src ports  \\
 & Uniq. dst ports  \\
 & Uniq. flows  \\ \hline
 & Num. of events per src IP addr  \\
 & Num. of events per dst IP addr  \\
 & Num. of events per src port  \\
 & Num. of events per dst port  \\
Rate & Num. of uniq. flows per src IP addr  \\
(10 features) & Num. of uniq. flows per dst IP addr  \\
 & Num. of uniq. flows per src port  \\
 & Num. of uniq. flows per dst port  \\
 & Uniq. src ports per src IP addr \\
 & Uniq. dst ports per dst IP addr \\ \hline
 
Top 10 hosts events & Num. of events from Top 10 src IP addrs \\
(2 features) & Num. of events to Top 10 dst IP addrs \\ \hline
Well-known 12 ports & Num. of events from/to port $x$ \\
(36 features) & Uniq. src IP addrs from/to port $x$ \\
 & Uniq. dst IP addrs from/to port $x$ \\ \hline

 & \texttt{gini}(num. of events,  src IP addr) \\
 & \texttt{gini}(num. of events,  dst IP addr) \\
 & \texttt{gini}(num. of uniq. src ports, src IP addr) \\
Gini-index ($x$, $y$) for & \texttt{gini}(num. of uniq. dst ports, dst IP addr)\\
for $x$, with respect to $y$ & \texttt{gini}(num. of uniq. dst ports, src IP addr)\\
(8 features) & \texttt{gini}(num. of uniq. src ports, dst IP addr)\\
 & \texttt{gini}(num. of uniq. src ports, dst port) \\
 & \texttt{gini}(num. of uniq. dst ports, src port) \\ \hline

\end{tabular}
}\label{tab:feature_list}
\end{table}

Although there are more sophisticated features that effectively represent security conditions, they are dependent on the semantics of specific events and the existence of specific types of security MBs. The feature set we define here is purposely  simple and general, because \name is designed to be generic and deployable at different enterprise networks having diverse security MBs.

Some of these features we define for \name are commonly used for other purposes such as real-time network monitoring and debugging. However, a significant and critical difference is that \name extracts features from the security event logs instead of network traffic (e.g., NetFlows or \texttt{pcap}).  Since security events are observed and reported by pre-defined rules, the feature values are not neutral in nature.  For example, a hundred TCP flows within ten minutes do not necessarily mean something wrong or bad by itself; whereas a hundred blocked TCP flows by a firewall does imply something suspicious if the usually observed number is, say, less than ten. Furthermore, the appropriate threshold to decide on the severity of the events is dependent on multiple factors, such as the capacity of network, the number of hosts and users, the number of hosted services, etc. For a large network, events corresponding to a few tens of blocked connections might be common; but if this number suddenly increases by an order of magnitude, an analyst might want to investigate further.  These differing characteristics of SIEM logs make it challenging to achieve good accuracy by naively employing traditional anomaly detection solutions. We show how \name overcomes this challenge with the contextual features and LVN in our evaluation in Section~\ref{sec:evalution}.

\section{Modeling phase: Unsupervised one-class modeling}\label{sec:modeling}
We describe the different aspects of modeling phase in this section. The common notations used are listed in Table~\ref{tab:notations}. 

\subsection{Training data sanitization}\label{sec:modeling:filtering}

The extracted features are used for training anomaly detection models. However, without ground truth it is not possible to separate out the more suspicious or anomalous data points from the rest of the data. As mentioned earlier, it is also practically challenging to  obtain ground truth for all event logs.
Therefore the data, i.e., the SIEM event logs, are likely to contain anomalies.  To deal with this limitation, \name filters out statistical outliers from the event logs before training a model for anomaly detection. More specifically, \name removes each feature's outliers based on the interquartile range (IQR); this is an efficient and effective way to remove statistically outlying points, although inevitably some anomalies would still remain within the data. The steps for filtering the outliers are presented in Algorithm~\ref{alg:iqr}. 
$\mathcal D$ denotes the dataset; 
with a slight abuse of notation, we use $\mathcal D_f$ to denote the column of values corresponding to a feature $f \in \mathbf f$. 
The filtering task is carried out in the inner \textbf{for} loop, with line~\ref{line:IQR-set-mean} setting the values of all outlier data points to the mean value of the dataset (for each corresponding feature).
In our experiments (Section~\ref{sec:evalution}), the parameter $\theta$ is set to 1.5 to compute the boundaries of feature values.

\subsection{Anomaly detection models}\label{sec:modeling:models} \name builds and utilizes three anomaly detection models for risk scoring, contextual analysis and low feature value identification and normalization.

\begin{table}[t]
\centering
\caption{Table of notations}
{
\begin{tabular}{ll} \toprule
Symbol & Description \\  \midrule
$\mathbf{B}$         & Set of middleboxes; $b$ denotes a middlebox \\
$\mathbf{C}$ & Context set; $c$ denotes a context \\
$\mathbf{f}$         & Feature vector, represented as a set \\
$\mathcal{D}$        & Dataset; $\mathcal{D}_f$ is the column corresponding to feature $f$  \\
$\mathbf{S}$ & Set of select data points and their anomaly scores\\ \bottomrule
\end{tabular}\label{tab:notations}
}
\end{table}

\begin{algorithm}[t]
\begin{algorithmic}[1]
\State \algorithmicrequire{~$\mathcal{D}$} \Comment{training set}
\For{$f \in \mathbf{f}$} \Comment{for each feature}
\State $q_3 \gets \texttt{percentile}(\mathcal{D}_{f}, 0.75)$
\State $q_1 \gets \texttt{percentile}(\mathcal{D}_{f}, 0.25)$
\State \text{iqr} $\gets q_3 - q_1$
\State \text{upp} $\gets q_3 + (\text{iqr} \times \theta)$
\State \text{low} $\gets q_1 - (\text{iqr} \times \theta)$
\State $m \gets \texttt{mean}(\mathcal{D}_{f})$
    \For {$\mathcal{D}_{f}[i] \in \mathcal{D}_{f}$} \Comment{change outlier values to $m$}
        \State $\mathcal{D}_{f}[i] \gets m \text{ if }  \mathcal{D}_{f}[i] \notin [\text{low}, \text{upp}]$ \label{line:IQR-set-mean}
    \EndFor
\EndFor
\State return $\mathcal{D}_f$
\caption{IQR-based Outlier filtering}
\label{alg:iqr}
\end{algorithmic}
\end{algorithm}

\begin{itemize}
    \item \textbf{MB models}: \name trains an anomaly detection model for each MB with all the features extracted in different event contexts; i.e., if there are $m$ MBs in an enterprise network, there are $m$ MB models trained.  The set of feature values at one time-slot is considered as one data point; therefore we expect an MB model to learn the correlation between the features of that MB within each specific time-slot.
    \item \textbf{Context models}: We build an anomaly detection model for each event context with only  the feature set belonging to that particular context (defined in Section~\ref{sec:CF:context}). This essentially translates to having multiple context models per MB. The context models enable \name to estimate and thereby explain the anomalies in specific contexts independently.
    \item \textbf{Per-feature models}: 
    A per-feature model is trained on a single feature; i.e., if there are $n$ features that \name extracts from across all middleboxes, then there are $n$ individual models corresponding to those features. For each model, we store the point corresponding to the lowest anomaly score while training; intuitively, this is a normal inlier point. This list of (potentially) normal feature values and scores are later used for low-value normalization (LVN), explained in Section~\ref{sec:scoring:lvn}. 
\end{itemize}

\begin{figure}[t]
    \centering
    \subfloat[clusters in per-feature model\label{fig:model-a}]{
        \centering
        \includegraphics[width=0.22\textwidth]{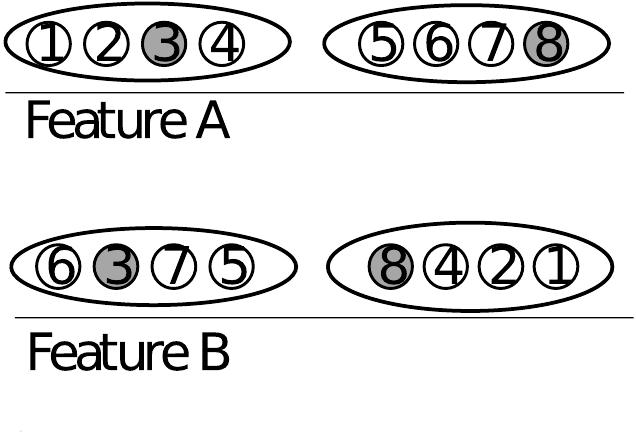}} \\
        \vspace{0.2cm}
    \subfloat[clusters in multi-feature model\label{fig:model-b}]{
        \includegraphics[width=0.30\textwidth]{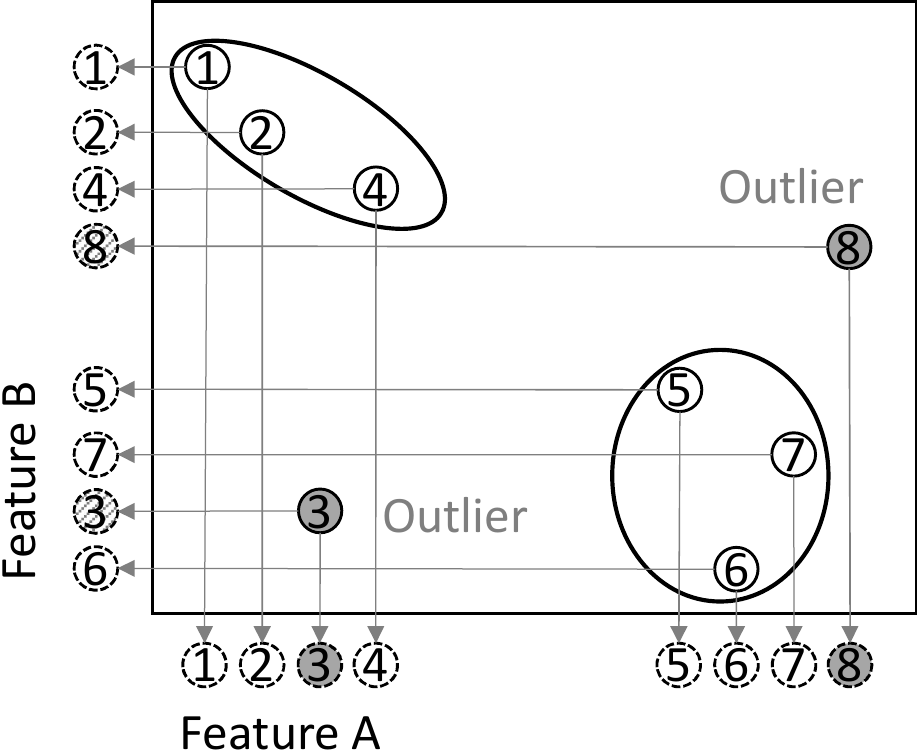}}
    \caption{(a) Per-feature modeling vs (b) Multi-feature modeling}
    \label{fig:local-global-compare}
    \vspace{0.3cm}
\end{figure}

We now discuss the pros and cons of different models. We use a simple example as illustrated in Fig.~\ref{fig:local-global-compare}. The left side of the figure shows two per-feature models for features A and B, both representing the same set of eight points (but along different dimensions). In both models, all points form two clusters of four points each; specifically, there is no point considered as outlier by these models. On the right is an MB model that also models the same eight points, but now using both the features A and B together, along the two axes. Now we see two clusters of normal inlier points, and two outlier datapoints~3 and~8. These two points are normal when modeled independently, but due to their correlation, they are seen as outliers by the MB model. Thus, the MB model will be able to detect outliers unseen by the per-feature model. 

One can also come up with a counter example, where the MB model represents a normal point as an outlier; such a scenario is possible when the normal point has (likely) independent feature values that are much lower than the majority of the points clustered by the MB model. 
This is one key characterstics of data points related to security events, in contrast to, say sensor readings. A very low sensor reading can be as anomalous as a very high sensor reading; but often, a relatively lower count of security events is likely a safer state of the network than a higher count. 
This is in particular true for those features for which lower values indicate normal state of the network. We refer to them as {\em absolute} features. For example, \textit{`number of active (open) TCP connections to a hosted server'} in the enterprise network is one such feature; a lower value indicates low number of client visits, while a relatively higher count could be due to an attack activity (or due to a genuine rise in the clients, which nevertheless is a statistical anomaly). In contrast are what we refer to as {\em relative} features. For example, the feature \textit{`number of events per source port'} can have a low value during a scanning activity; whereas, it can also have a low value during a normal time when there are less sources (and hence less source ports) initiating connections. Therefore, lower value of this feature does not help in coming to a conclusion on what is happening in the network. Later in Section~\ref{sec:scoring:lvn}, we adopt LVN to automatically allow \name to learn this specific characteristics of security events.

Another shortcoming of an MB model is that it is not able to explain or provide the context of the anomaly. An MB model barely gives any information for the detected anomalies; but the per-feature model on the other hand tells us exactly which feature caused the anomaly. Similarly, the context model also gives the set of contextual features that triggered the anomaly, thereby straightaway informing the context of the anomaly. \name supplements the shortcoming of an MB model with the context models.  
The anomaly score from the context models reports the most relevant context for the given time-slot by measuring the anomaly score only with the features in each specific context.  Thus specifying the context, \name drastically reduces the number of feature that an analyst needs to investigate. 

\subsection{Modeling pipeline}

\begin{algorithm}[t]
\begin{algorithmic}[1]
\State \algorithmicrequire{~$\mathbf{B}, \mathcal{D}, \mathbf{C}$} \Comment{$\mathbf{B}$:MBs, $\mathcal{D}$:training set, $\mathbf{C}$: context set}
\For{$b \in \mathbf{B}$} \Comment{for each MB}
    \State $\mathcal{D}^{b} \gets \texttt{IQR\_filter}(\mathcal{D}^{b})$\label{line:TR:IQR-filter}
    \State $\mathcal{M}^{\text{mb}}_{b} \gets \texttt{train}(\mathcal{D}^{b})$ \Comment{MB model}\label{line:TR:mb-model}
    \For{$c \in \mathbf{C}$}
        \State $\mathcal{M}^{\text{context}}_{b,c} \gets \texttt{train}(\mathcal{D}^{b, c})$ \Comment{context models}\label{line:TR:context-model}
    \EndFor
    \For{$f \in \mathbf{F}$}
        \If{\texttt{is\_absolute\_feature}($\mathcal D^b_f$)} 
            \State $\mathcal{M}^{\text{per-feature}}_{b,f} \gets \texttt{train}(\mathcal D^b_f)$ \Comment{per-feature models}
            \State $\mathbf{S}^{\text{per-feature}}_{f} \gets \texttt{anomaly\_score}(\mathcal{M}^{\text{per-feature}}_{b,f}, \mathbf{D}^b_f)$ \label{line:TR:pf-model}
            \State $v^{f}_{\min} \gets \texttt{value\_at\_min\_score}(\mathbf{S}^{\text{per-feature}}_{f})$
            \State $\mathbf{L}^{b}_{f} \gets v_{\min}$ \Comment{min value and corresponding point}\label{line:TR:lvn}
        \EndIf
    \EndFor
\EndFor
\State return $\mathcal{M}^{\text{mb}}, \mathcal{M}^{\text{context}}, \mathbf{L}$
\caption{\name training pipeline}
\label{alg:modeling_phase}
\end{algorithmic}
\end{algorithm}

We summarize the modeling pipeline in 
Algorithm~\ref{alg:modeling_phase}. $\mathcal D^{b,c}$ denotes the dataset of feature values corresponding to an MB $b \in \mathbf B$ and context $c \in \mathbf C$. 
The algorithm starts by removing the outliers from the dataset (line~\ref{line:TR:IQR-filter}) since the training set possibly contains anomalies. %
For each MB, line~\ref{line:TR:mb-model} trains an MB model $\mathcal{M}^{\text{mb}}_{b}$ for MB $b$. 
Similarly, for each MB and each context, line~\ref{line:TR:context-model} trains a context model $\mathcal{M}^{\text{context}}_{b,c}$.  Within the second \textbf{for} loop, we train per-feature model for each {\em absolute} feature (line~\ref{line:TR:pf-model}), and use that model to compute the score of each point being an anomaly. Since lower scores indicate normal points, we store the data points corresponding to the lowest scores in line~\ref{line:TR:lvn}, to use in the analysis phase for low-value normalization. 

\section{Analysis Phase}\label{sec:scoring}
In the analysis phase, \name computes the anomaly scores for each measurement time-slot, e.g., for every 10-minute interval, using MB models and context models. While the scores from the MB models is used for the final ranking of the time-slots for the anomalous behavior (described in Section~\ref{sec:scoring:agg}), the scores from context models are used to explain the anomalies (Section~\ref{sec:scoring:investigation}). The ranking along with the context level information assists a SOC analyst to prioritize and better understand the anomalies, and to carry out further investigations. 
First, we motivate the need for low-value normalization, and how \name achieves this.
\vspace{-0.2cm}
\subsection{Low-value normalization (LVN)}\label{sec:scoring:lvn}

\begin{figure}[t]
    \centering
     \subfloat[before applying LVN\label{fig:lvn-a}]{
        \includegraphics[width=0.23\textwidth]{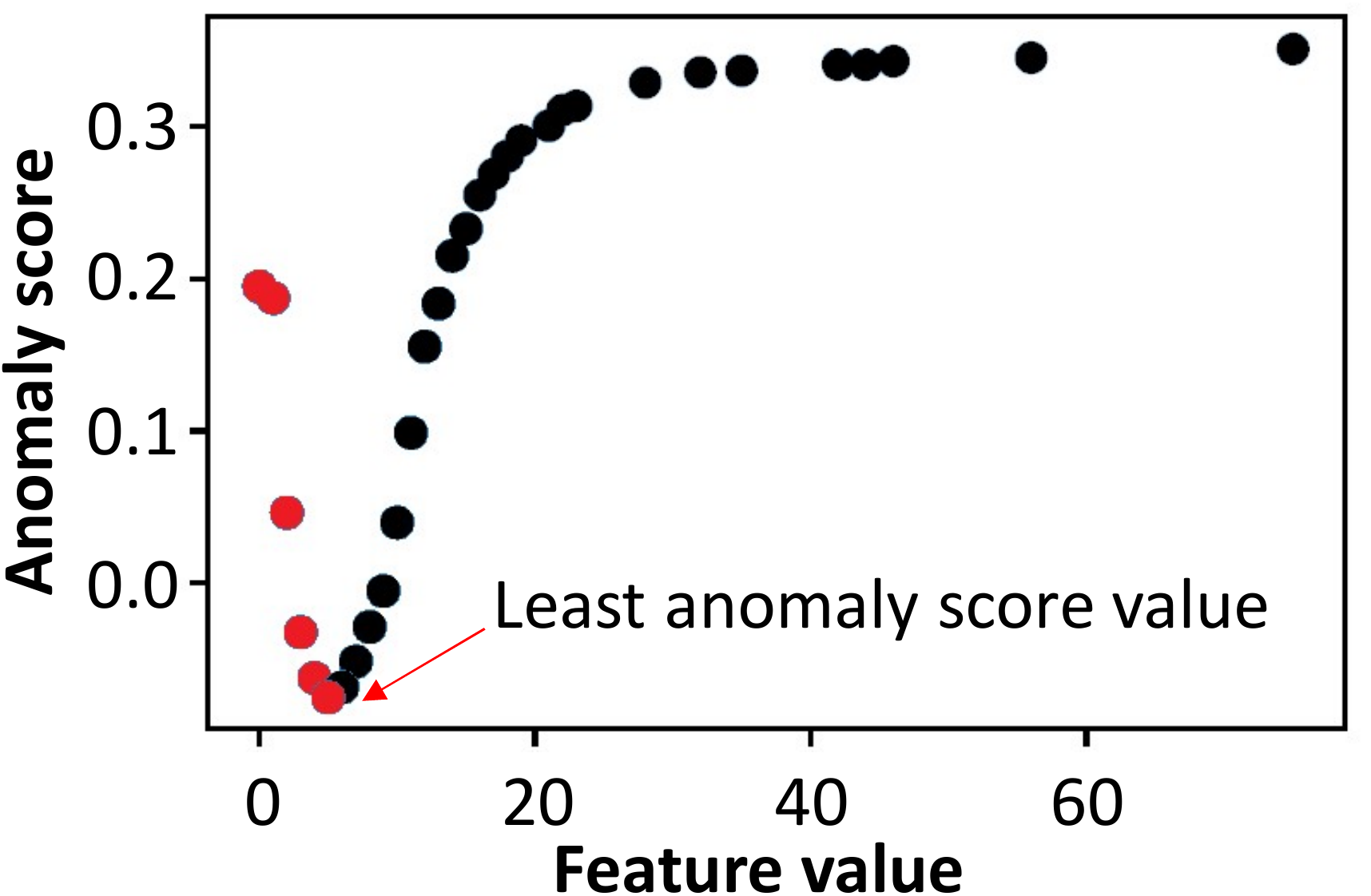}}
    \subfloat[after applying LVN\label{fig:lvn-b}]{
        \includegraphics[width=0.23\textwidth]{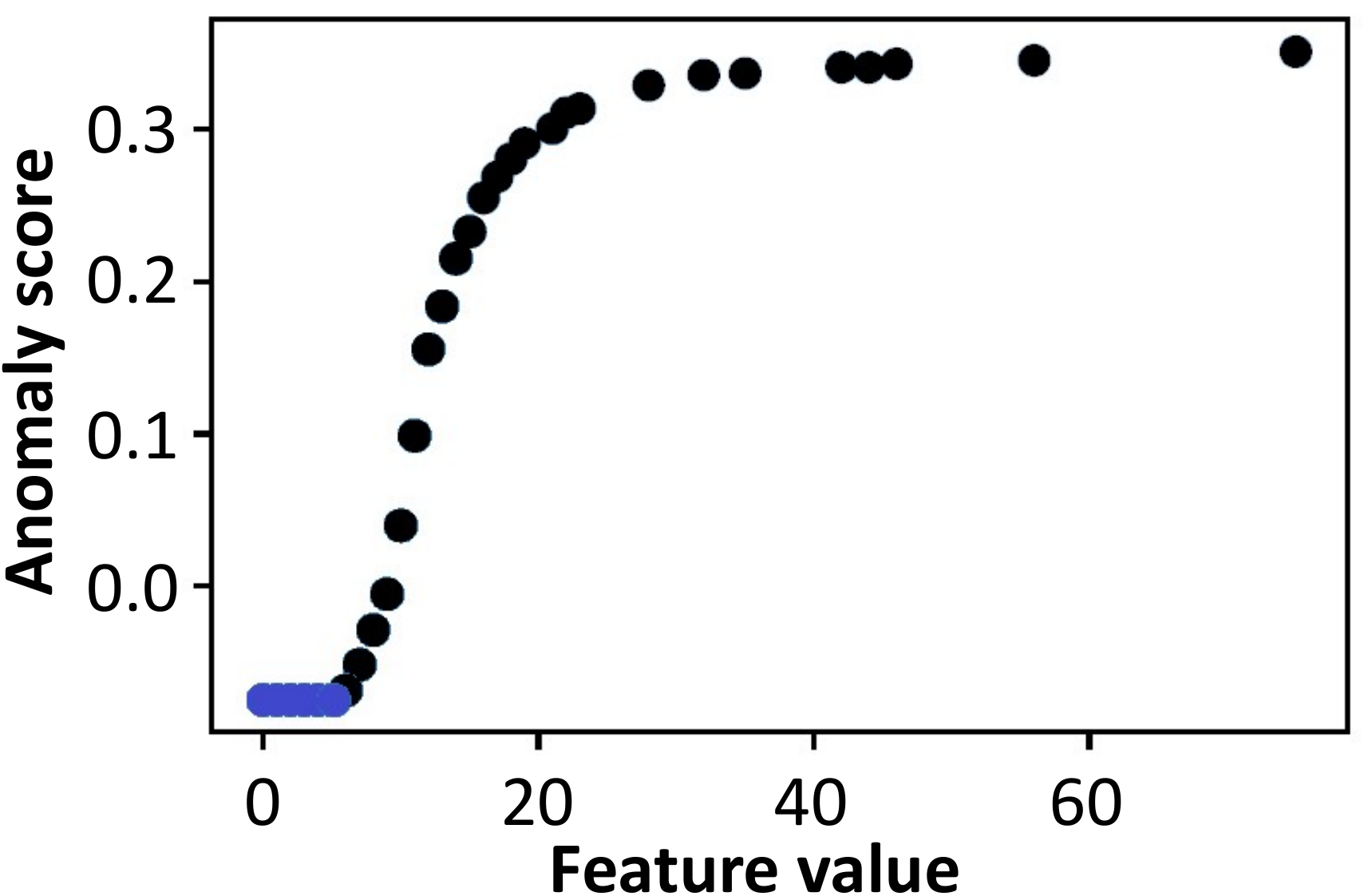}}
    \caption{Illustration of false positives due to low values, and anomaly score flattening using low-value normalization.}
    \label{fig:lvn}
\end{figure}

Fig.~\ref{fig:lvn-a} illustrates the problem of falsely detecting low values of {\em absolute} feature (that are in red dots) as outliers. This problem is exacerbated when models are trained with numerous features, as is in our solution framework.  We solve the problem by normalizing the data in the lower range of the feature values. Fig.~\ref{fig:lvn-b} illustrates our goal. To carry out normalization, we need to estimate and store the feature values that correspond to the lowest anomaly scores from the per-feature models.  As explained earlier, \name does this in the training phase (refer lines~\ref{line:TR:pf-model}-\ref{line:TR:lvn} in Algorithm~\ref{alg:modeling_phase}). Thus we obtain a set of low values, say $v^f_{\min}$ for each {\em absolute} feature $f \in \mathbf{F}$. Subsequently, in the analysis phase (see Section~\ref{subsec:analysis-pipeline}), \name uses these minimum score values to modify the data such that, each data point for an {\em absolute} feature $f$ is not smaller than $v^f_{\min}$.

\subsection{Time-slot scoring and ranking}\label{sec:scoring:agg}
One major goal of our work is to rank the anomalies in SIEM event logs; since events are processed in time-slots, the goal is essentially to rank the time-slots. As motivated earlier, such a ranking would allow an analyst to prioritize the events for further investigation. 

\name uses the MB models to estimate anomaly scores for time-slots. Thus, each MB model provides one score. If there are $\tau$ time-slots for analysis, the range of the ranks according to the scores is $[1, \tau]$. Let $r^b_t$ denote the rank of a time-slot $t$ obtained from an MB model $\mathcal{M}^{\text{mb}}_{b}$ based on the features from middlebox $b$. 
Furthermore, if there are $k$ MBs deployed at an enterprise network, there will be $k$ ranks for each time-slot, as each MB ranks each time-slot independently. One approach to aggregate the $k$ scores for a time-slot would be to compute the Euclidean distance of the rank vector, $[r^1_t, r^2_t, ... , r^k_t]$ for time-slot $t$, from the origin $[0,0, ... ,0]$, and then take the nearest distance as the highest ranked anomaly. However, such a method may rank a time-slot with moderate ranks from all middleboxes (example~2 in Table~\ref{tab:dist_example}) as more anomalous than a time-slot with high scores from not all middleboxes (example~3 in Table~\ref{tab:dist_example}). Whereas, in practice the contrary is preferred, as one or two middleboxes indicating a high probability of anomaly should not be ignored.

To overcome the above problem, \name uses the percentile of the ranks for computing the aggregate score for a time-slot. Percentile is also a better metric than rank in our context, as the number of time-slots may change depending on the enterprise, the network size, the number of users, etc.  We denote $\rho^b_t$ as the percentile corresponding to rank $r^b_t$ (from MB $b$, for time-slot $t$). The third column in Table~\ref{tab:dist_example} gives the percentile of example ranks in the first column. 

Based on the percentiles for a time-slot $t$ from all MBs, we compute the aggregate score as a distance:
\begin{equation}
\label{eq:distance}
    d_{t} = \sqrt{\sum_{b\in \mathbf{B}}{((\rho^{b}_{t})^{2})}}
\end{equation}
The higher the distance, the more anomalous is the time-slot. The last column in Table~\ref{tab:dist_example} gives the percentile-based distance measures for the rank tuples in the first column. With this measure, the third example is more anomalous than the second example, as opposed to the naive distance computation presented in the second column. 

\begin{table}[t]
\centering
\caption{Examples illustrating the aggregation anomaly scoring based on ranks and percentiles}
{
\begin{tabular}{c|rrcc} \toprule
\# & ($r^{1}_{t}$,$r^{2}_{t}$,$r^{3}_{t}$,$r^{4}_{t}$) & naive & ($\rho^{1}_{t}$,$\rho^{2}_{t}$,$\rho^{3}_{t}$,$\rho^{4}_{t}$) & $d_{t}$ \\ \midrule 
1. & (1, 1, 1, 1) & 2.00 & (0.99, 0.99,0.99, 0.99)  & 1.98 \\ %\hline
2. & (60, 50, 50, 20) & 94.87 & (0.40, 0.50, 0.50, 0.80) & 1.14 \\ %\hline
3. & (1, 90, 1, 90) & 127.28 & (0.99, 0.10, 0.99, 0.10)  & 1.41 \\ % \hline
4. & (100, 100, 100, 100) & 200.00 & (0.00, 0.00, 0.00, 0.00)  & 0.00 \\ \bottomrule
\end{tabular}\label{tab:dist_example}
}
\vspace{0.3cm}
\end{table}

\subsection{Analysis pipeline}
\label{subsec:analysis-pipeline}

% Analysis phase
\begin{algorithm}[t]
\begin{algorithmic}[1]
\State \algorithmicrequire{~$\mathbf{B},  \mathbf{C}, \mathcal{M}^{\text{mb}}, \mathcal{M}^{\text{context}}, \mathcal{D},\mathbf{L}$} 
\For{$b \in \mathbf{B}$}
    % LVN
    \For{$f \in \mathbf{f}$}
        \If{\texttt{is\_absolute\_feature}($f$)}
            \For{$\mathcal{D}_{f}[i] \in \mathcal{D}_{f}$}
                \State $\mathcal{D}_f[i] \gets \texttt{max}(\mathbf{L}^{b}_{f}, \mathcal{D}_f[i])$ \Comment{LVN} \label{line:TS:lvn}
            \EndFor
        \EndIf
    \EndFor
    \State $\mathbf{S}^{\text{mb}} \gets \texttt{anomaly\_score}(\mathcal{M}^{\text{mb}}_{b}, \mathcal{D}^{b})$ \Comment{MB score} \label{line:TS:mb-score}
    \State $\mathcal{R}^{\text{mb}}_{b} \gets \texttt{rank\_percentile}(\mathbf{S}^{\text{mb}})$\label{line:TS:mb-rank}
    
    \For{$c \in \mathbf{C}$} \label{line:TS:context-loop}
        \State $\mathbf{S}^{\text{context}_{c}} \gets \texttt{anomaly\_score}(\mathcal{M}^{\text{context}}_{b,c}, \mathcal{D}^{b}_{c})$ \Comment{context score} \label{line:TS:context-score}
        \State $\mathcal{R}^{\text{context}}_{b,c} \gets \texttt{rank\_percentile}(\mathbf{S}^{\text{context}_{c}})$  \label{line:TS:agg-context}
    \EndFor
\EndFor
\State $\mathcal{R}^{\text{agg}} \gets \texttt{distance\_rank}(\mathcal{R}^{\text{mb}})$
\State return $\mathcal{R}^{agg}$, $\mathcal{R}^{\text{context}}$ \Comment{aggregated ranks} \label{line:TS:agg-mb}
\caption{Analysis phase}
\label{alg:analysis_phase}
\end{algorithmic}
\end{algorithm}

Algorithm~\ref{alg:analysis_phase} presents the pipeline of the analysis phase. Prior to estimating the anomaly scores of data points (time-slots) in the test set, \name normalizes the data points with values lower than the lowest observed value in the training set ($\mathbf{L}$ is the set of lowest-score data point for each feature); note, this is performed only for the absolute features (line~\ref{line:TS:lvn}). In other words, the inner {\bf for} loop performs LVN. Next, \name computes two sets of anomaly scores and percentiles of all data points (time-slots). The first set of anomaly scores are obtained from the MB models, and their percentiles computed subsequently (line~\ref{line:TS:mb-score}-\ref{line:TS:mb-rank}). The second set of scores and percentiles for the time-slots are computed by the context models in the last {\bf for} loop (lines~\ref{line:TS:context-loop}), to obtain the scores based on each of the contexts defined earlier (Section~\ref{sec:CF:context}). The penultimate line calls the function \texttt{distance\_rank} which computes the aggregate score based on Equation~\ref{eq:distance}, and ranks the time-slots such that the time-slot with the highest distance score becomes the top-ranked anomalous point.

\subsection{Investigation pipeline}\label{sec:scoring:investigation}
\begin{figure}[t]
    \centering
    \includegraphics[width=0.44\textwidth]{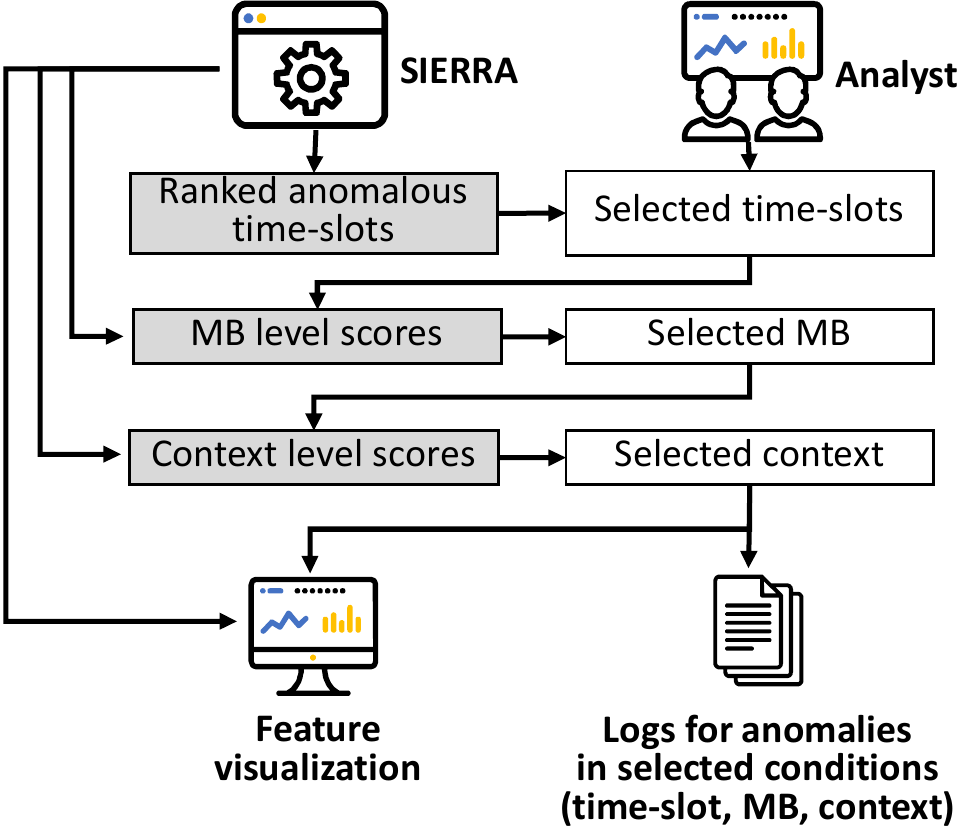}
    \caption{Investigation pipeline for an analyst}
    \label{fig:analyst-flow}
\end{figure}

Fig.~\ref{fig:analyst-flow} illustrates an investigation flow SIEM log events assisted by \name. At the beginning, \name provides the ranked anomalous time-slots to a security analyst. The security analyst may decide to further investigate some of the top-ranked anomalies within the testing period considered. For selected anomalous time-slot selected by the analyst, \name provides the prioritized list of MBs and contexts that are most relevant, and these are based on the scores obtained from the MB models and context models. Finally, \name also provides a time series representation of the most important context features that lead to the time-slot being ranked as anomalous. We explain this further with real case studies in Section~\ref{sec:scoring:case}. This prioritized information helps in explaining the anomalous points to the analyst.

\section{Performance Evaluation}\label{sec:evalution}

\subsection{Experiment setup}\label{sec:dataset}

\noindent \textbf{Enterprise SIEM event log dataset.}
To evaluate our system, we collected over 29 millions of SIEM events logged during the period Mar.~2020 to Jul.~2020 from a reputed commercial cyber security company. The logs come from three different security middleboxes of an enterprise network; see Table~\ref{tab:siem-data} for details.\\

\noindent \textbf{Ethical concerns.}
Since events logged by middleboxes could potentially reveal sensitive information of the concerned enterprise, strong security measures were put in place when it came to accessing the data. No identifiable information (such as customer name) was present in the data. A small number of selected employees of the security company (including some of the co-authors) were given access to process the data purely for research purposes.  The data was stored within the network of the security company; and therefore, the data processing, model training and testing have to be carried out within the same network. Thus, data did not leave the security company's network and premise. Respecting privacy concerns, we also do not reveal the exact firewalls and IDS in use; but they are from well-known solutions from different vendors.\\

\begin{table}[t]
\centering
\caption{Enterprise SIEM event logs dataset for evaluation ($\text{FW}_{1}$, $\text{IDS}_{1}$ and $\text{FW}_{2}$ are from three different vendors) \label{tab:siem-data}}

{
\begin{tabular}{ccc} \toprule
Notation &  Num. of events &  Unique hosts \\ \midrule
$\text{FW}_{1}$ &   15232 K &  76 K \\
$\text{IDS}_{1}$  & 7820 K &  3 K\\
$\text{FW}_{2}$  &  6349 K & 33\\
Total & 29401 K &  78 K \\ \bottomrule
\end{tabular}
}
\end{table}

\noindent \textbf{Ground truth.}
The ground truth on the dataset of SIEM events was obtained via multiple means. First and foremost, the investigations carried out by the analysts of the security company marked some of the events (and time-slots) as indicative of security incidents (breaches, attacks, etc.). Second, multiple threat intelligence sources (including VirusTotal~\cite{virus2021}) were used to verify part of the anomalies in the dataset; these resulted in detecting some characteristics in the logs corresponding to a botnet campaign. Third, rules as well as heuristics to verify inter-feature consistency were used for additional detection of events related to malicious activities. Finally, manual analysis was also carried out to determine the presence of security incidents. Nevertheless, we do not make a claim that each event in the dataset is accurately labeled; however, we took care to ensure the validity of the results reported here. 
Indeed, the metrics used for evaluation in this work (below) are helpful in such scenarios; they are also supportive of the possibility that the dataset may not being exhaustively and perfectly labeled. Henceforth, we refer to the time-slots with security incidents, botnet campaigns and other malicious activities as {\em positive time-slots}. \\

\noindent \textbf{Performance metric.} To evaluate the performance of \name, we use \textit{precision at rank n}, or ($P@n$)~\cite{moffat2008rank}, as the metric. $P@n$ measures the fraction of correctly detected positive data points (time-slots) within the top $n$ ranked time-slots. 
With such a definition, note that, if $m$ of the top-$n$ ranked data points are false positives, there are also $m$ false negatives that did not come up in the ranking due to the false positives occupying those ranks. $P@n$ is a widely used performance metric in information retrieval systems~\cite{moffat2008rank} as well as in anomaly detection systems~\cite{zhao2019lscp}, where precision at high ranks are more important than the overall precision.

In a SIEM anomaly detection system, $P@n$ represents the capability to detect the known positive cases (i.e., security incidents) within a given analysis budget $n$, which often indicates the available resources at a SOC, e.g., the number of budgeted man-hours for security analysis and investigations.  $P@n$ is defined as:
\begin{equation}\label{eq:accuracy}
P@n = \frac{T(n)}{n},
\end{equation}
 $T(n)$ being the number of positive time-slots within the top $n$  ranks.

\label{subsec:ex-config}

To evaluate suitability of the models for anomaly detection, we consider five models with distinct approaches. i)~LOF (Local Outlier Factor)~\cite{breunig2000lof} is useful for detecting local outliers of a dense cluster, in particular when the clusters have different densities, by considering both distance and density in feature space. This characteristic works negatively when the data points are biased, but a small difference does not necessarily mean an outlier. ii)~iForest (Isolation Forest)~\cite{liu2008isolation} is computationally efficient and robust to the varying distances of the outliers, unlike distance and density-based approaches such as LOF. iii)~OCSVM (One-class Support Vector Machine)~\cite{scholkopf2001estimating} is expected to give accurate anomaly scores in multi-variate feature space where scoring a feature value as outlier is dependent on other features.  
iv)~VAE (Variational Autoencoders)~\cite{kingma2019introduction} is a neural network that learns a distribution-based latent representation of the input data, and uses the reconstruction error of the test data (based on the learned latent representation) to detect anomalies. 
Lastly, v)~LSCP (Locally Selective Combination in Parallel Outlier Ensembles)~\cite{zhao2019lscp} is an anomaly detection system which uses an ensemble of multiple local detectors. In our experiments, we use 40~LOFs as the local detectors of LSCP.

We employ the above one-class models as such in the naive form for comparison with \name. In other words, to analyze the effect of \name, we adopt the one-class models within \name framework and compare with the naive implementations of these models. 
In other words, within \name framework, the models will be supported by LVN and features will be segregated based on context and MB, as multiple context models and MB models are used in \name, with the MB models used for anomaly scoring.\\

\noindent \textbf{Feature extraction.}
The models applying the contextual features (CF) are trained with 62 features defined in Table~\ref{tab:feature_list}, but extracted from 30, 31 and 6 different contexts corresponding to $\text{FW}_{1}$, $\text{IDS}_{1}$, and $\text{FW}_{2}$, respectively. 
For instance, we have a total of 1,860 features ($62 \text{ features} \times 30 \text{ contexts}$) for $\text{FW}_{1}$.  Thus, a feature, `\texttt{number of unique destination host}' is engineered as thirty features for different contexts, like, `number of unique destination hosts when the event is \{\texttt{Warn/Info}\} severity, the source host is in \{\texttt{LAN/WAN/DMZ}\}, the destination host is in \{\texttt{LAN/WAN/DMZ/Others/Exception}\}, and the flow is for \{\texttt{Request/Response/Others}\}' to/from the server.

The experiments are conducted as follows. We split the entire dataset into weeks; there are  19 (full) weeks in total. Consider a week as a unit of rolling window. The first run of an experiment consists of a model trained on the first 8~weeks and tested on the following one week; then we roll the window by one week, and train again using the next 8-week (week~2 to week~9) of data and test on the new and following $10^\text{th}$ week of data. Continuing in this fashion, we execute the experiment 11 times, where the $11^\text{th}$ experiment trains on data from week~11 to week~18, and tests on week~19. We report the average of these 11 experiments as results. 
The length of the time-slots for feature extraction and modeling is set to 10 minutes. For the metric $P@n$, the value of $n$ is set to 100 in our experiments, unless otherwise stated.

\vspace{-0.2cm}

\subsection{Component Evaluation}

\noindent \textbf{Baseline setting.} We now evaluate different baseline models for \name. For all these baseline models, we use the feature set defined in Table~\ref{tab:feature_list} from all the event logs of all MBs, but without considering the context of the events and LVN techniques. Upon applying the contextual features (CF) in \name, we aim to achieve detection capability as good as the naive use of well-known OC-models with the feature set, while additionally providing contextual information with the detection.\\

\begin{figure}[t]
    \centering
    \includegraphics[width=0.46\textwidth]{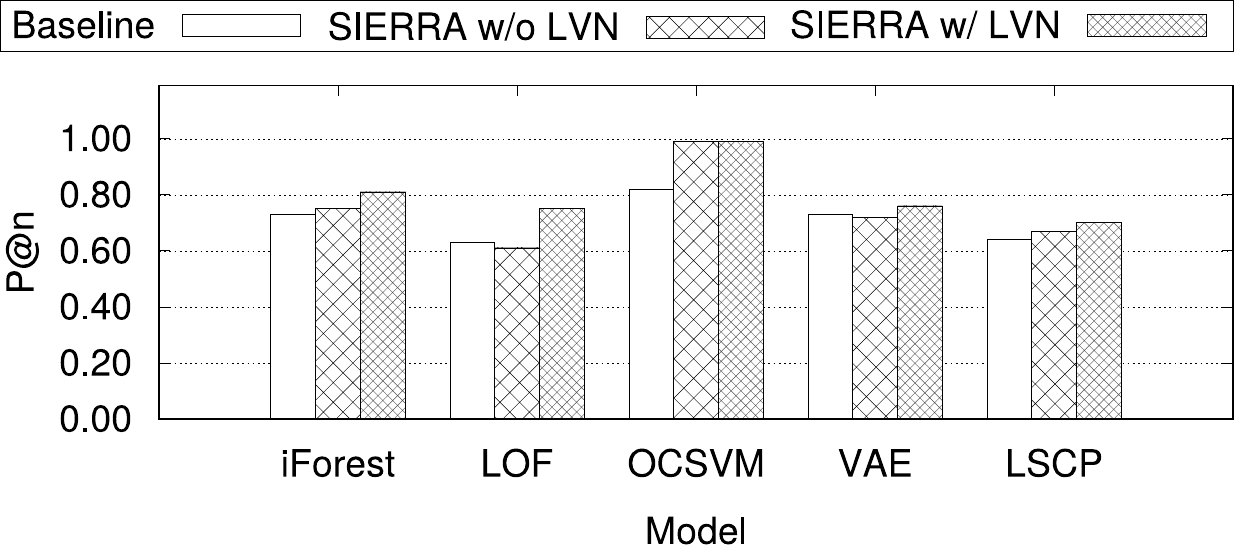}
    \caption{Comparing \name with the corresponding baseline anomaly detection implementations}
    \label{fig:acc_global}
\end{figure}

\begin{figure}[ht]
    \centering
    \subfloat[$P@n$ of MB model for $\text{FW}_1$\label{fig:acc-fw1}] {
        \includegraphics[width=0.46\textwidth]{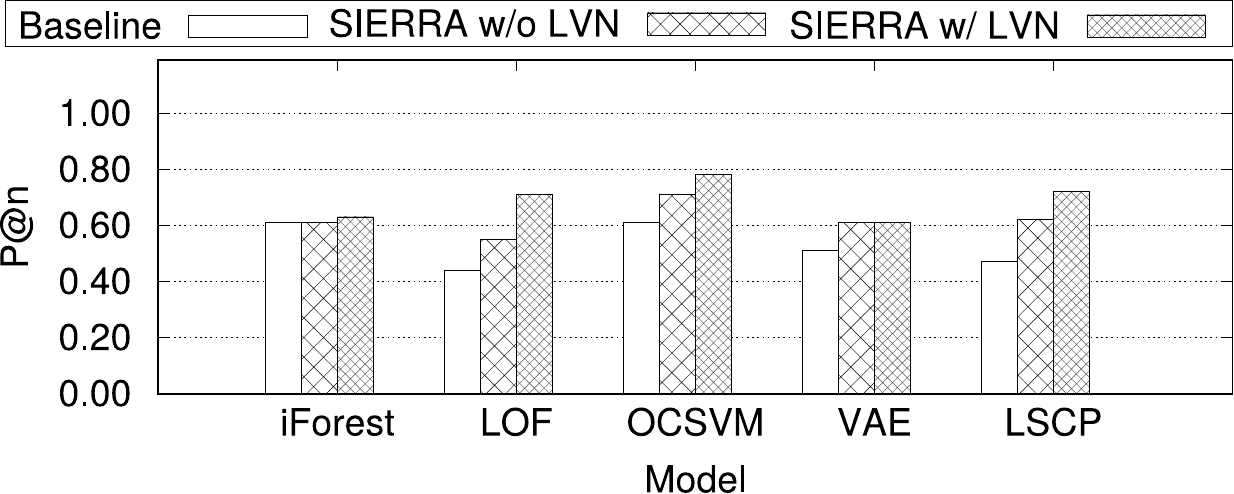} }
        \\
    \subfloat[$P@n$ of MB model for $\text{IDS}_1$\label{fig:acc-ids1}] {
        \includegraphics[width=0.46\textwidth]{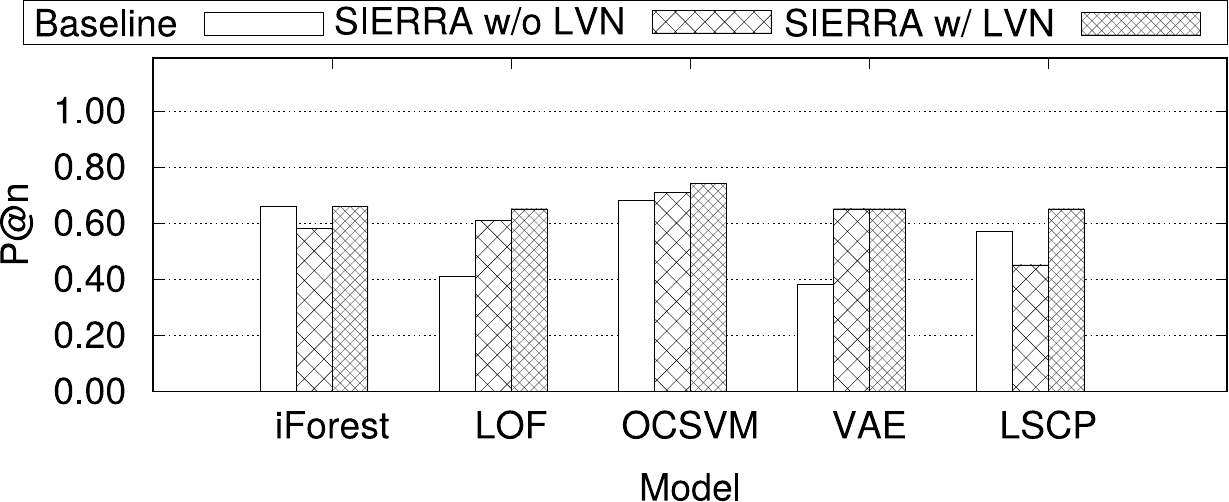} }
        \\
    \subfloat[$P@n$ of MB model for $\text{FW}_2$ \label{fig:acc-fw2}]{
        \includegraphics[width=0.46\textwidth]{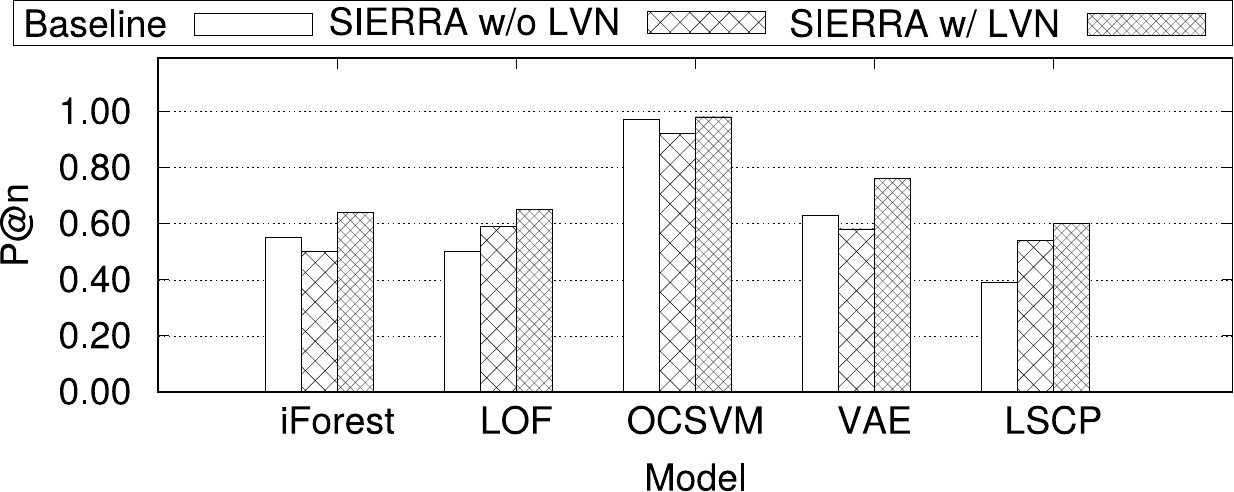}}
    \caption{Comparing \name with the corresponding baseline anomaly detection implementations for individual MBs}\label{fig:cf_lvn_benefit}
\end{figure}

\noindent \textbf{Performance improvement due to Contextual Features and LVN.}
Fig.~\ref{fig:acc_global} presents the results of the overall detection capability after aggregating the scores from all MBs with percentile-based scoring. And Fig.~\ref{fig:cf_lvn_benefit} shows the results for the individual MBs.
Needless to say, the top-$n$ time-slots are different for the different MBs (in the ground truth data), as each middlebox has its own rules for triggering events. 
Thus, each individual MB's performance in Fig.~\ref{fig:cf_lvn_benefit} is measured with its own ground truth.
Whereas, for evaluating the performance of aggregating scores from all MBs (Fig.~\ref{fig:acc_global}), the ground truth is formed by taking union of the positive labels in all of three MBs.\\
We make multiple observations:

i)~The adoption of LVN on the contextual features clearly helps in improving the precision for the top-$n$ ranking ($n = 100$).

ii)~The best performance improvement due to \name and the highest $P@n$ are achieved with OCSVM. $P@n$ for \name is 17\% point higher than the corresponding baseline OCSVM. 

iii)~Finally, we also observe that, the contextual features on its own does not always lead to performance improvement. This is not surprising as contextual features are defined and segregated for each context, and thus have smaller and less diverse values.  This in turn affects the model when the feature values are in the lower end of the spectrum.  And that is the reason why the performance improves when we combine LVN with contextual features --- LVN is designed to precisely deal with false positives due to low-value anomalies.  
For example, consider the {\em absolute} feature `number of unique destination hosts' that ranges from zero to a few hundreds.  When we extract this feature for each context separately, the unique number of hosts is measured in a specific condition, like, `number of unique destination hosts \textit{when the destination host is a server in DMZ and the source host is an external client}'.  The contextual feature `number of unique server in DMZ with external clients' has low and a short range of values (less than three) in our dataset (recall, feature values are extracted for each time-slot).  Short range values easily tend to have a skewed distribution, and the model may assign a high anomaly score to a rarely appearing smaller value. Whereas, the corresponding naive feature without the context condition has a smoother distribution and gives a smaller anomaly score to the value with a small difference.

However, LVN normalizes the scores of low-value points and solves this issue. The performances of iForest in Figures~\ref{fig:acc_global} and~\ref{fig:cf_lvn_benefit} depict this. While contextual features of \name lead to a reduced precision in comparison to the baseline models, adopting LVN increases the precision beyond the baseline models.\\

\begin{figure}[ht]
    \centering
    \subfloat[Recall for first test window\label{fig:recall-3024}]{
    \includegraphics[width=0.44\textwidth]{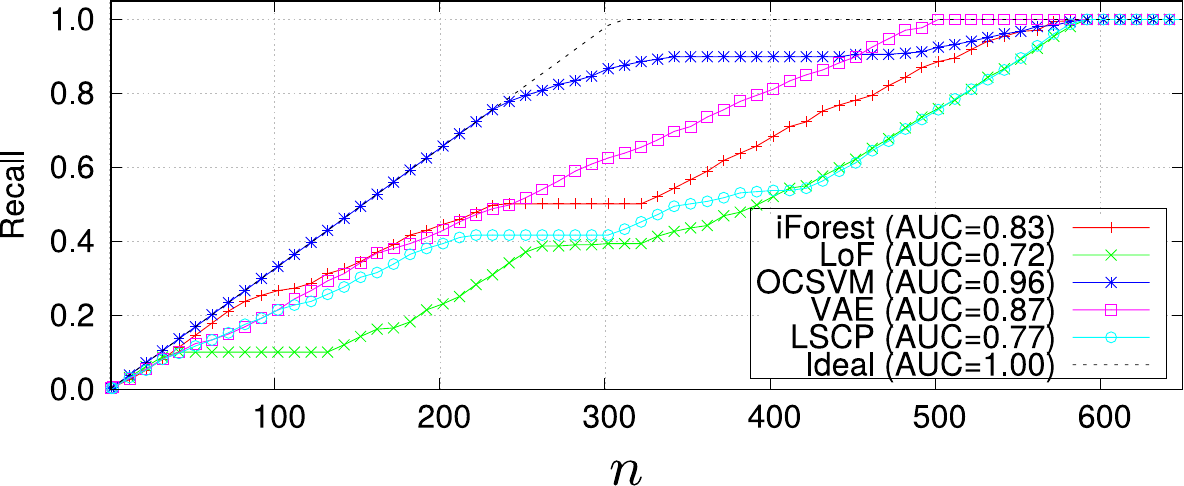}}
    \\
    \subfloat[Recall for second test window\label{fig:recall-4032}]{
    \includegraphics[width=0.44\textwidth]{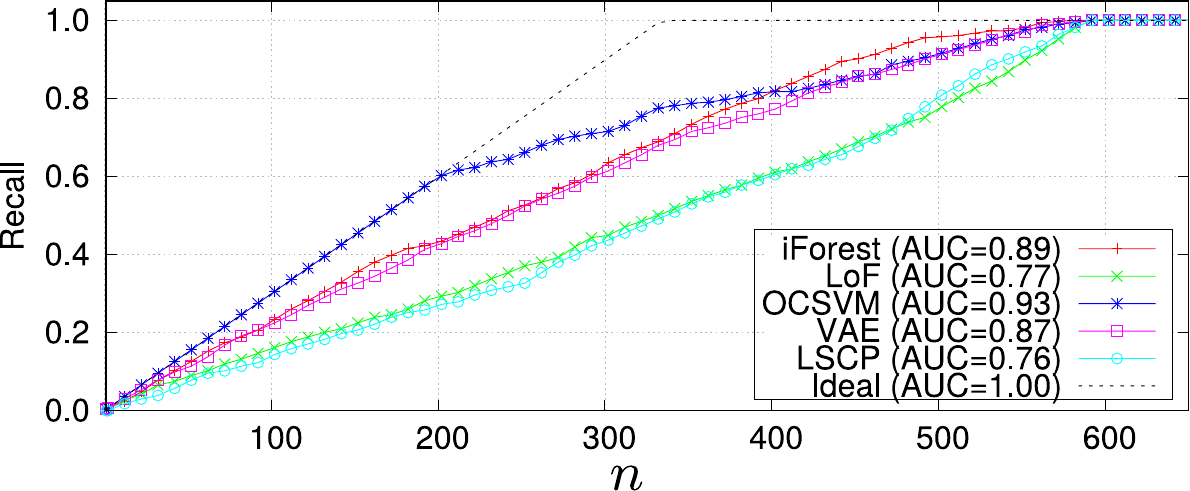}}
    \\
    \subfloat[Recall for third test window\label{fig:recall-5040}]{
    \includegraphics[width=0.44\textwidth]{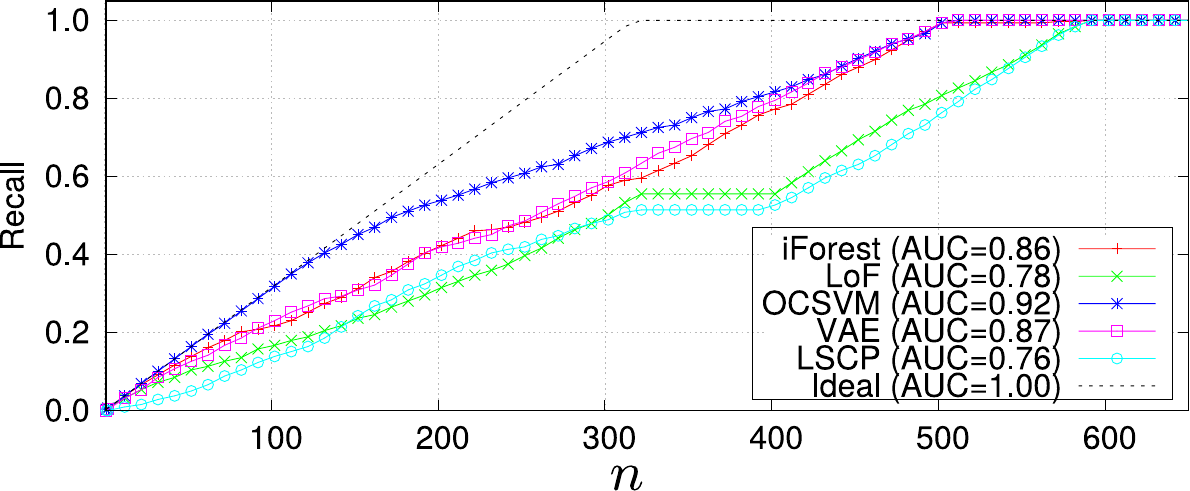}}
    \caption{Model comparisons for varying budgets ($n$)
    }\label{fig:recall-cdf}
\end{figure}

\noindent \textbf{Varying the budget}. 
Next we analyze the performance for varying budget for SOC analysis, i.e., $n$.  Fig.~\ref{fig:recall-cdf} plots the recall (Y-axis) for different values of $n$ (X-axis).  We show the results for the first three moving test windows (weeks). There are more than 300 positive time-slots (i.e., those with verified security incidents) in these three weeks. The {\em ideal} line plots the maximum that can be detected within the top-$n$ ranked time-slots; thus the comparison of a model should be with this ideal line. From the figures, we observe that \name with OCSVM still performs the best. In the first time window, \name with OCSVM captures 266 of all anomalous points in the top 300 time-slots; that is, it achieves a recall of 0.86 while the other models have recall around 0.6 or less.  The area under the curve (AUC) is also highest for OCSVM for all three time windows.\\

\noindent \textbf{Analysis effort for top-n time-slots.}  
\name provides the ranked anomalous time-slots for prioritizing investigation; the number of time-slots is configurable based on the resources available for analysis.  \name helps in reducing the investigation efforts by specifying the event context and top-related features. Analysis of 300 time-slots for a week (as is in our experiments) translates to around 42 time-slots per day, which is a reasonably low number of time-slots for a large organization to investigate.\\

\subsection{Overhead analysis}
\noindent \textbf{Experiment configuration.}
We analyze the computational time for training \name models. We measure the elapsed time for training the context models ($\mathcal{M}^{\text{context}}$) and middlebox models ($\mathcal{M}^{\text{mb}}$) for the experiments described in Section \ref{sec:dataset}. The experiments runs on a machine with four physical Intel(R) Xeon(R) CPU E5-2620~v3 processors and 256GB~RAM, equipped with 2.40GHz Tesla~P4 GPU with 8GB~RAM, and running on a Ubuntu 16.04 OS. \name is implemented with Python~3.8 and Tensorflow~2.5.0 to be compatible with GPU driver and CUDA versions. \\

\noindent \textbf{Training and testing overhead.}
The average times to train $\mathcal{M}^{\text{context}}$ (for all contexts) over 11 experiments are approximately \textbf{10}, \textbf{3}, \textbf{5} minutes, for $\text{FW}_{1}$, $\text{IDS}_{1}$, and $\text{FW}_{2}$, respectively. 
Similarly, for $\text{FW}_{1}$, $\text{IDS}_{1}$, and $\text{FW}_{2}$,
the average times to train one $\mathcal{M}^{\text{mb}}$ are \textbf{20}, \textbf{15} and \textbf{5} minutes, respectively.

The time gap between MBs are mainly due to the number of valid contexts as well as the diversity and distribution of values. In the worst-case scenario where all contexts have random values, the training time for $\mathcal{M}^{\text{mb}}$ is approximately 3~hours. 
In practice, we assume models might require re-training only once in a month or twice; thus even a worst-case training time of a few hours is acceptable. As expected, testing is quite fast; it takes only less than 5, 4, and 2 seconds to test on one week's of data  (i.e., 1008 data points) from $\text{FW}_{1}$, $\text{IDS}_{1}$, and $\text{FW}_{2}$.

\subsection{Comparison with state-of-the-art systems}

We compare \name with one of the state-of-the-art anomaly detection systems, \textit{DeepLog} proposed by Du et al.~\cite{du2017deeplog}.
DeepLog is a log modeling system utilizing Long Short-Term Memory (LSTM) without prior knowledge of each event type and log source. In contrast to the solutions that are dependent on specific log types~\cite{yen2013beehive,oprea2018made,najafi2019malrank}, DeepLog aims to be generic anomaly detection solution without the domain knowledge of the target network and the semantic of each log type~\cite{du2017deeplog}; \name too has the same goal.\\

\noindent \textbf{DeepLog configuration.} DeepLog has two detection methods, one each for detecting execution path anomaly and parameter value anomaly.  Since our SIEM log dataset does not have execution logs, we compare with the latter.  In the parameter value modeling, 
DeepLog models time series of the values for each unique event type.

Most of the IDS and firewall logs have IP addresses and port numbers as values, but neither of them are suitable to model the numerical fluctuation.  Therefore, though it is not defined in DeepLog design, we set three parameter configurations, i) \textbf{naive}: the number of events for each event type, ii) \textbf{feature}: the feature set used in \name for each event type, iii) \textbf{context feature (CF)}: the feature set for each event context. Note that a context is not always a super set of event types because an MB may define a general event type without regarding one or multiple context properties of events, e.g., \texttt{`WARN:TCP\_BLOCKED'} for both network direction.  We use 144 time-slots (one day in 10min time-slot size) as the sequence input for LSTM.  Following the DeepLog design, we normalize the values using the mean and standard deviation of all values from the training set, and use the last one week of the training set as the validation set to build the scoring model based on the prediction (refer~\cite{du2017deeplog} for details). To rank the time-slots, we take the highest anomaly score from the event type models in naive and feature configurations and from the context models in CF configuration.\\

\noindent \textbf{\name configuration.} We use OCSVM for all three anomaly models of \name, i.e., per-feature, context and MB models. 
To be fair with the scoring mechanism of DeepLog, we evaluate the performance of \name using MB score. \\

\begin{figure}[t]
    \centering
    \includegraphics[width=0.44\textwidth]{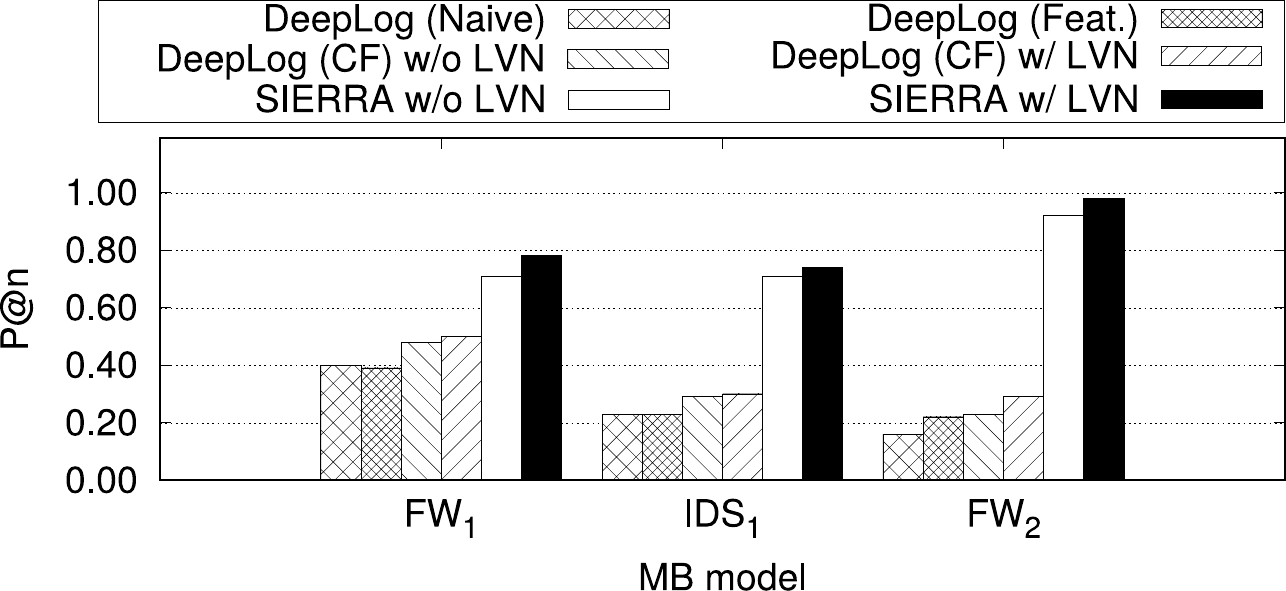}
    \caption{Comparison of \name and DeepLog~\cite{du2017deeplog}}
    \label{fig:acc_deeplog}
\end{figure}

\noindent \textbf{Results.} Fig.~\ref{fig:acc_deeplog} shows the performance of both \name and DeepLog. The figure demonstrates a significant advantage of \name; while DeepLog achieves 0.4 or less precision (at top 100 time-slots) for the three MBs, \name achieves around 0.8 for $\text{FW}_{1}$ and $\text{IDS}_{1}$, and close to 1.0 $P@n$ for $\text{FW}_{2}$.
In the three experiment configurations for DeepLog and adoption of LVN on CF configuration, we observe insignificant difference between the configurations since they share the following reasons:

i)~Anomalies defined by multiple features: One critical reason for the low performance is that many anomalies in the ground truth are not detectable by only counting the events (which is used as feature in DeepLog). While \name models a feature set in a multi-dimensional model, DeepLog assumes only one parameter stream for an LSTM model which leads to multiple isolated models for each event type (naive) or a feature (feat. and CF).

ii)~Sequential dependency: Even for the same testing value, predictions vary widely depending on the prior sequence of values.  For example, after a high spike or sudden fall of a feature, a following value in a normal range may be predicted as anomaly as it is not an expected sequence. While the outlying values deviating from prior values tend to have higher anomaly score, the values in the regular range may also result in high anomaly score leading to false positives in top 100 ranks.

DeepLog also predicts high anomaly score for low-value data points.  From our analysis and understanding of the model, this is not only because the low-value points are themselves rare, but because the sequences having a low-value data point in a random position are rare and diverse to build a reliable sequential model.  We observe a little advantage by adopting LVN in DeepLog as shown in Fig.~\ref{fig:acc_deeplog} (DeepLog (CF) w/ LVN); yet the improvement is clearly smaller than that of \name, primarily because DeepLog performs single-feature sequential modeling.

\section{Case study}\label{sec:scoring:case}

In this section, we illustrate the capability of \name in aiding the investigation of real anomalies detected on the test set.  We investigate anomalies step-by-step starting from one outstanding anomalous time-slot.

The analysis phase of \name generates ranked anomalies from all middleboxes; in our dataset, we have three MBs. From the list, we pick a top-ranked time-slot and an MB to investigate the activities captured by the MB. 
In this case study, $12105^\text{th}$ time-slot from the fifth test window is a clear anomaly at MB-level scores. Of the three MBs, the MB that produces the highest score in this particular time-slot is $\text{FW}_1$, and the context with the highest score is \texttt{[Warn,WAN-DMZ,Request]}.  This context is interpreted as \textit{`blocking events at $\text{FW}_1$ triggered by flows originating from the hosts in WAN and directed to DMZ servers in the target network'}.\\

\begin{figure}[t]
    \centering
    \subfloat[Anomaly scores from context model \texttt{[Warn,WAN-DMZ,Request]}.\label{fig:tp-a}]{
        \includegraphics[width=0.48\textwidth]{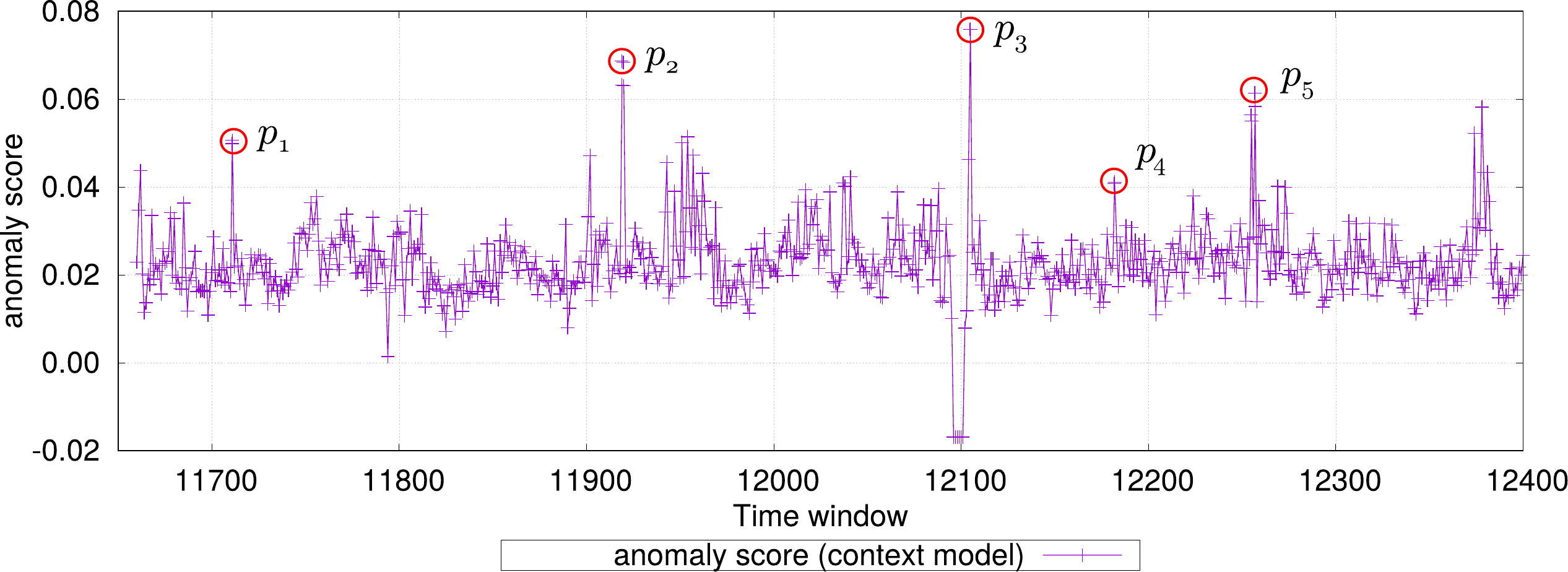}}
        \\
    \subfloat[The number of all events, of the events to top 10 internal hosts and of the events from top 10 external hosts\label{fig:tp-b}]{
        \includegraphics[width=0.48\textwidth]{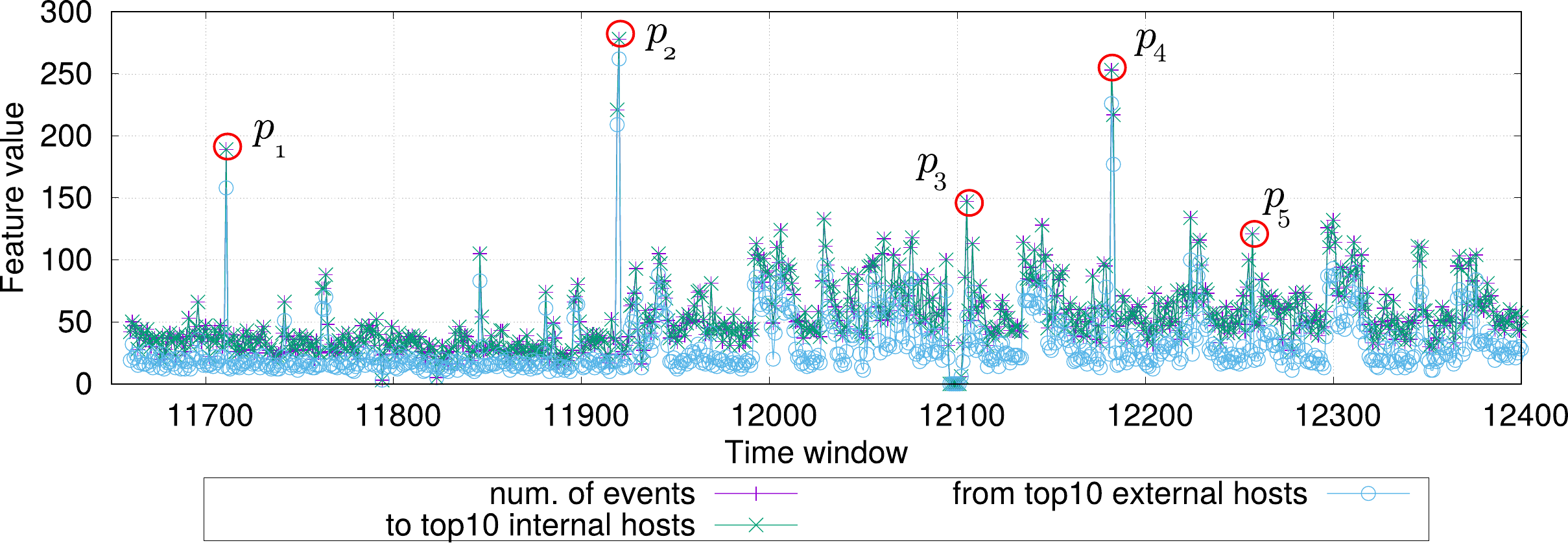}}
        \\
    \subfloat[The number of unique internal ports, external ports, and incoming flows\label{fig:tp-c}]{  
        \includegraphics[width=0.48\textwidth]{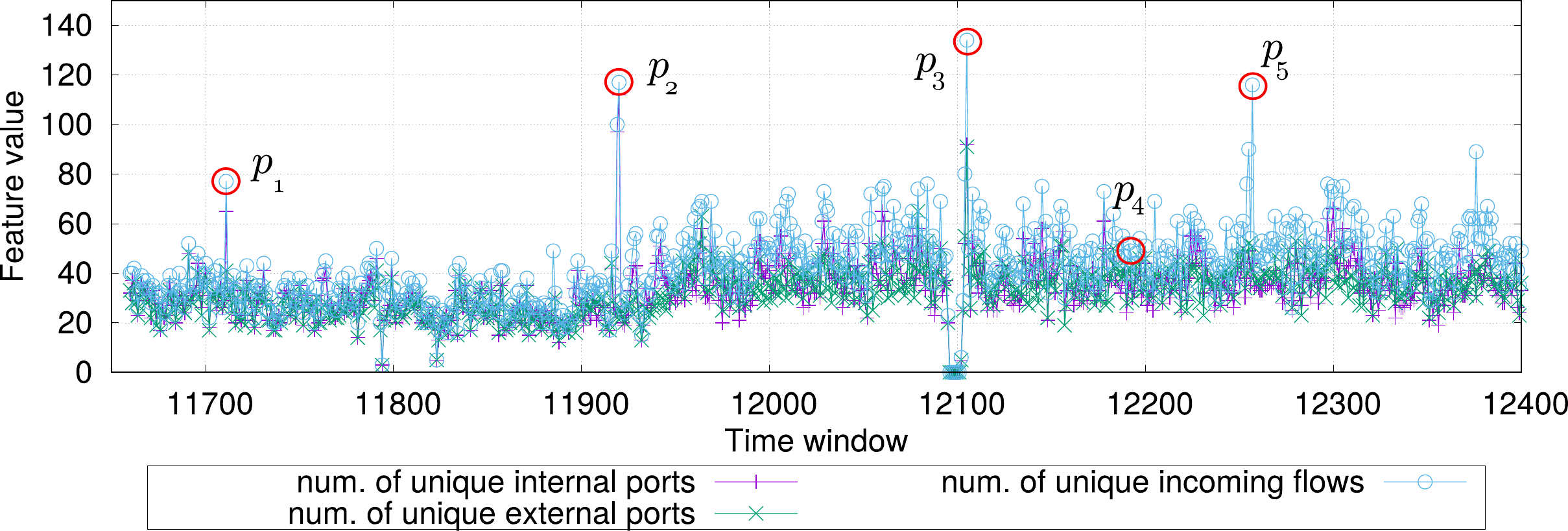}}
    \caption{Anomaly scores and feature values in sample time-slots [11650:12400]. We highlight five sample cases $p_{1}, ... p_{5}$ from \texttt{[Warn,WAN-DMZ,Request]} context model of $\text{FW}_1$.}
    \label{fig:tp-fortinet-score}
\end{figure}

\noindent \textbf{Target time-slots}. In the given context and condition, that is, $b=\text{FW}_1$ and $c= \texttt{[Warn,WAN-DMZ,Request]}$, \name provides visual representation of the anomaly scores and features to an analyst. Fig.~\ref{fig:tp-fortinet-score} shows what an analyst sees for investigating the anomalous time-slots.  Out of the tens of features in the above context, \name automatically selected the 10 most relevant features for these anomalies based on the variance during the time period.  For ease of discussion, we select six of these features for further analysis.  Besides the anomaly scores, we plot the feature values in the neighbouring time-slots, i.e., [11650:12400] time-slots. We observe anomalies not only in the time-series of the anomaly scores (Fig.~\ref{fig:tp-a}), but also from time-series plots of the features values (Figures~\ref{fig:tp-b} and~\ref{fig:tp-c}).  Motivated by the multiple sharp spikes observed in the time-series of the feature values,  we include four additional time-slots that have spikes in Fig.~\ref{fig:tp-b} and \ref{fig:tp-c}); these are depicted as $p_{1}, \dots,  p_{5}$ (including our initial target time-slot, $p_{3}$).\\
%on the figures. 

\noindent \textbf{Contextual feature analysis.} One observation from Figures~\ref{fig:tp-b} and~\ref{fig:tp-c} is that the five points have anomalously high values in different features. For example, the sharp increase in the anomaly score (Fig.~\ref{fig:tp-a}) at the time-slot $p_{2}$ comes with high values in the `number of events' (Fig.~\ref{fig:tp-b}). Meanwhile, $p_{3}$, which has the highest anomaly score in Fig.~\ref{fig:tp-a},  has high `number of unique incoming flows' (Fig.~\ref{fig:tp-b}). 
This difference is not surprising because different types of intrusions may be captured by different sets of features, e.g., port scanning having numerous unique destination ports, whereas web attacks target only web application ports (e.g., 80/443).  This observation supports the motivation of \name, i.e., defining an anomaly needs to consider the environment dependent conditional correlations among numbers of features.

\begin{table}[ht]
\centering
\caption{Activities observed in investigation}
{
\begin{tabular}{lcccc} \toprule
Activity & Src hosts & Src ports & Target hosts & Target ports \\ \midrule
Act. 1 & fixed & ranged & fixed & 443 \\ 
Act. 2 & random & random & fixed & random \\
Act. 3 & fixed & fixed & fixed & 23 \\ 
Act. 4 & fixed & fixed & fixed & random \\ 
Act. 5 & subnets & ranged & fixed & random \\ \bottomrule
\end{tabular}
}\label{tab:activity-analysis}
\end{table}

\addtolength{\tabcolsep}{-2pt}
\begin{table}[ht]
\centering
\caption{Rank percentile in context-level scores}
{
\begin{tabular}{ccccccc} \toprule
 Case & Context $\rho$ & Act. 1 & Act. 2& Act. 3 & Act. 4 & Act. 5 \\ \midrule
 $p_{1}$ & 99\% & H & M & - & - & - \\ 
                     $p_{2}$ & 99\% & H & L & - & - & - \\ 
                     $p_{3}$ & 99\% & L & H & - & - & - \\ 
                     $p_{4}$ & 97\% & M & M & H & - & - \\
                     $p_{5}$ & 99\% & - & M & - & H & M \\ \bottomrule
\end{tabular}
}\label{tab:case-analysis}
\end{table}
\addtolength{\tabcolsep}{2pt}   

\noindent \textbf{Activity analysis}. As the last step of analysis, we look into the events in the target context in each time-slot and group the events sharing common characteristics into activities. The five activities thus identified are listed in Table~\ref{tab:activity-analysis}; they explain these anomalies. High density of Act.~1 leads to high number of unique source ports without a corresponding increase in the number of unique source IP addresses and destination ports. In contrast, rise in Act.~4 increases only the number of unique destination ports.  We observe presence of one or more activities in all target time-slots but in different densities, i.e., the number of events grouped into the activity, - L (low), M (medium), and H (high); these are presented in Table~\ref{tab:case-analysis}. While there are always a certain level of activities in a network as background noises, e.g., low level of Act.~2, rise of new activities and change of activity density are captured by the \name features even in the cases when multiple activities co-occur. From the context \texttt{[Warn,WAN-DMZ,Request]} and density of activities, we classify Act.~1 and Act.~3 to attacks targeting specific internal HTTPS and Telnet servers, respectively. Whereas, Act.~2 and Act.~4 show typical scanning behavior, attempting to connect to multiple ports. 
Act.~5 is from a set of known (external) network prefixes that are part of a massive scanning campaign; they target web servers on non-standard ports of IoT devices and home-routers.

\section{Related works}\label{sec:relatedwork}

Many research works analyze specific log or data sources for detecting specific attacks, e.g., 
network traffic~\cite{NADA-2018, GEE-CNS-2019}, 
DNS data~\cite{bilge2011exposure,ahmed2019monitoring,sun2020detecting,khalil2016discovering}, web-proxy logs~\cite{lee2020building, yen2013beehive,oprea2018made,oprea2015detection,najafi2019malrank}, etc.
DNS-based solutions typically detect malicious activities by capturing inherent behaviors (e.g., periodic communication and DGA) of malware. Use of web-proxy logs in addition gives visibility to URLs and HTTP payloads. Yen et al. proposed Beehive~\cite{yen2013beehive} for suspicious activity detection based on web-proxy logs; its main contribution is in addressing the inconsistency and incompleteness of log data.  Another proposal MADE~\cite{oprea2018made} aims at detecting malicious network activities with features extracted from a web-proxy. MalRank~\cite{najafi2019malrank} attempts to classify the malicious communications by building a knowledge graph based on web-proxy, DNS and DHCP logs as well as from external information sources, such as, open-source intelligence (OSINT) and cyber threat intelligence. 
Bartos et al.~\cite{bartos2016optimized} propose a
malware detection system using supervised learning and based on HTTP connections. To be robust to variants, the proposed system aims to learn invariant properties as features from web-proxy logs.

However, these systems targeting detection of malware communications and malicious activities have one or more of the following disadvantages: dependence on one specific (MB) log, detection on only targeted behaviors (e.g., periodic communication or DGA), and use of labels to define the malicious activities. Another serious limitation is that, DNS and HTTP logs do not cover the behaviors between internal hosts, that is a critical threat in an enterprise; similarly incoming attacks from the Internet are also not captured by these logs. \name, in comparison, focuses to detect any forms of anomalies, e.g., unseen internal communications, increase of long-lived connections, etc., that are unlikely detected by the above models.

Studies on host-based logs have mainly focused on abstraction of logs and causal-analysis~\cite{kwon2018mci,zeng2021watson,liu2018towards}, in particular, against Advanced Persistent Threat (APT) attacks~\cite{milajerdi2019holmes,hassan2020tactical,alsaheel2021atlas}. While fine-grained host logs provide causal relationship between events, they are also very noisy. Thus abstracting and tracking relevant events relies on seed knowledge (e.g., a known incident) from an expert or synthetic dataset; an approach different from our motivation and design. 

Anomaly detection systems using an unsupervised approach largely relax the downsides of the supervised approaches that require the labeled dataset. However, building an anomaly detection model often comes with assumptions that are difficult to be guaranteed in SIEM environments. 

The early systems modeling log-sequences~\cite{lou2010mining} and log counts of individual log type~\cite{xu2009detecting} require a fixed set of log types to define the log sequences and count the same type of logs; whereas a SIEM system has multiple MBs and periodically updates the logging policies. Studies overcoming the diversity of log formats and deviation in logging terminologies have proposed systems that adopt NLP models and deep learning techniques.  LogTransfer~\cite{chen2020logtransfer} vectorizes the words in the logs with GloVe~\cite{pennington2014glove}, and LogBERT~\cite{guo2021logbert} captures the contextual semantics of log sentences using Bidirectional Encoder Representations from Transformers (BERT)~\cite{devlin2018bert}. LogRobust~\cite{zhang2019robust} overcomes the instability issues due to change of log template, sequence, and changing logging statements, by extracting the word semantic of the logs with Bidirectional LSTM model. Adaptive Deep Log Anomaly Detector (ADA)~\cite{yuan2020ada} uses dynamic models and threshold selection with LSTM models to enhance accuracy. Autoencoders have also been widely adopted for anomaly detection.  VeLog proposed by Qian et al.~\cite{qian2020anomaly} achieves sequential modeling of execution paths and the number of execution times using variational autoencoders (VAE). Catillo et al. proposes AutoLog~\cite{catillo2022autolog} which models term-weightings with autoencoders.

Common shortcomings of systems focusing on the semantic anomalies are: i)~Assuming logging of normal operation which defines the normal conditions semantically. In particular, learning word semantics from data without regarding parameters has the disadvantage of being misled by the presence of words learned (e.g., an incoming \texttt{ACCEPT} event to a rare port might be abnormal, even while \texttt{ACCEPT} is learned as normal). ii)~Determining anomaly based on single log lines (or events) containing  anomalous terms or anomalous order of actions (this may be a fair assumption in system logs but not in SIEM logs).  For example, in a set of events with simple templates, \texttt{`TCP Denied IP * Port *'} and \texttt{`TCP Allowed IP * Port *'}, the sequence of events, word semantics, and a single log line hardly imply an anomaly.  In particular, the systems that regard only the word semantics can easily be misled by the appearance of \texttt{Denied} and \texttt{Allowed}, as they may be the only deviations in certain log entries.
Besides, LogAnomaly~\cite{meng2019loganomaly} proposes template2Vec technique to model synonyms and antonyms of words using log templates. Whereas DeepLog~\cite{du2017deeplog} models both the execution path and the parameter changes more than word and log appearance, by modeling the normal workflows from an arbitrary system without domain knowledge of keywords in the free-text logs.

Despite the flexibility of these sequence-based state-of-the-art systems, one limitation is that they assume sequences of normal events. While security MBs might log normal events (although) occasionally, it is not realistic to assume logging of {\em reliable sequences} of normal events. Unlike the systems where usual behaviors are dependent on services and users, SIEM systems have multiple external adversaries, making it unrealistic to expect sequential behaviors in both normal and abnormal scenarios.

\name considers the characteristics of SIEM logs that are not taken into account in the above anomaly detection systems, yet, on the other hand, uses the contextual features that are generally applicable to different types of security MBs.

\section{Discussions}
\noindent \textbf{Dependency on the quality of rule sets of MBs.}
At this point, it is also pertinent to ask, how dependent would \name be on the MBs, and in particular, on the quality of events generated by MBs. Clearly, any anomaly detection system is highly dependent on the visibility of activities happening in an enterprise network. That is, the activities have to be logged by at least one of the deployed MBs for an anomaly detection system to analyze the events. If rules are biased, the visibility of \name naturally follows the visibility. For example, if an MB is a web firewall that monitors network traffic only on the web ports (e.g., port 80 and 443), \name is only allowed to detect anomalies in the web activities. \\
The other problem MBs have is that, they might log unwanted or unnecessary information. Since these are triggered by rules and policies, the logging is also consistent; e.g., a (less useful) firewall rule to log all \texttt{ACCEPT} connections will result in events whenever new connections are accepted. As long as the number of new connections (per unit time) does not result in a sudden peak or valley, \name also would not detect it as an anomaly. Thus rules and policies leading to less informative events will not affect \name's performance.

In terms of ranking of anomalies, an overall increase of the feature values, as well as the event volumes caused by a sensitive and general rule set, may possibly affect \name in the training phase. \name may learn a less sensitive model against the anomalies belonging to the same context of the flooded events. On the other hand, an anomaly from a context with less background noise naturally achieves a higher score with a small deviation.
\\

\noindent \textbf{Potential weakness of \name.}
\name shares the common weakness of unsupervised learning systems.  Although \name includes training set filtering module, low-rate or long-term persistent attacks possibly influence the models during re-training and may make the system less sensitive.  While the worst-case scenario for \name is a training dataset with  random values, it is however challenging for an adversary to generate random noises because of the following reasons. First, the MBs log only events that are triggered by their rules/policies, and these rule sets are not disclosed to the adversaries. Second, an adversary hardly has global control over the other adversaries that may attack the network.

The anomalies that \name focuses on are from a set of events (essentially, their statistical features). Therefore, \name is unlikely to detect an anomaly caused by a single event, as well as anomalies in the values or words in an event (which has been the focus of many existing systems).  For example, one first-seen event, due to the first occurrence of an activity or a newly deployed MB rule, may not be ranked high  unless it leads to anomalous feature values in the context. But \name allows us to set a new context for the event types that need special consideration, e.g., `\texttt{critical}' in event severity property.\\

\noindent \textbf{Possibility of opportunistic attacker choosing less-active time slot.} This requires the attacker to have insider information of the interested enterprise network, to learn what is being logged and when. More importantly, note that an attacker does not get any feedback from the model, as the model output goes to the analyst.\\

\noindent \textbf{Reproducibility.}
One can reproduce \name with publicly available libraries.  All the ML algorithms (e.g., OCSVM, IsolationForest, etc.) used in anomaly models are available from Python's scikit-learn library~\cite{scikit-learn}. For feature extraction, we provide the full list of features in Table~\ref{tab:feature_list} in Section~\ref{sec:dataset}.  Due to the confidentiality and sensitive information in the dataset, unfortunately, we cannot disclose the dataset.

\section{Conclusions and future directions}\label{sec:conclusion}

In this work, we developed \name, that is able to rank anomalies in an enterprise network based on the event logs from multiple and different security middleboxes deployed. Our goal has been to develop a solution for practical deployment; hence \name employs unsupervised models, and provides the context and relevant features of the detected anomalies to support further investigation.

The evaluations on real logs from an enterprise show promising results. This is our first step towards solving the problem of dealing with large number of generated events at a SOC. 

As analysts investigate the top-ranked anomalies, they also generate ground truth information pertaining to these anomalies. While the investigated anomalies are only a fraction of all the events generated by a SIEM, and therefore is not exhaustive, it nevertheless generates a small but confident labeled dataset. Thus, building an anomaly detection system that utilizes the confident and labeled dataset with large sets of unlabeled data will be our next step.

\section*{Acknowledgment}
This research is supported by the National Research Foundation, Prime Minister’s Office, Singapore under its Corporate Laboratory@University Scheme, National University of Singapore, and Singapore Telecommunications Ltd. We thank the anonymous reviewers and the shepherd for helping to improve this work. Last but not the least, we thank the late Sanjeev Shankar (VP of Product Engineering, Trustwave), who helped us identify the problem and had been a constant source of inspiration for our research.

{\footnotesize \bibliographystyle{IEEEtran}
\bibliography{references}}
\newpage

\end{document}